\documentclass[fleqn,usenatbib]{mnras}

\usepackage{newtxtext,newtxmath}
\usepackage[T1]{fontenc}

\DeclareRobustCommand{\VAN}[3]{#2}
\let\VANthebibliography\thebibliography
\def\thebibliography{\DeclareRobustCommand{\VAN}[3]{##3}\VANthebibliography}

\usepackage{xcolor}
\usepackage{graphicx}	% Including figure files
\usepackage{amsmath}	% Advanced maths commands
\usepackage{pdflscape}  % To include landscape pages

\title[Flyby-induced warps in protoplanetary discs]{Scattered light signatures of flyby-induced warps in protoplanetary discs}

\author[Katie L. Milsom et al.]{
Katie L. Milsom,$^{1}$
Tim J. Harries,$^{1}$
C. J. Nixon,$^{2}$
\\
$^{1}$Department of Physics and Astronomy, University of Exeter, Stocker Rd, Exeter EX4 4QL, UK\\
$^{2}$School of Physics and Astronomy, University of Leeds, Sir William Henry Bragg Building, Woodhouse Ln., Leeds LS2 9JT, UK
}

\date{Accepted 2026 May 19. Received 2026 May 12; in original form 2025 November 26
}

\pubyear{2025}

\begin{document}
\label{firstpage}
\pagerange{\pageref{firstpage}--\pageref{lastpage}}
\maketitle

\begin{abstract}
 We explore the observational signatures of flybys in scattered light images of protostellar discs. The warps are modelled using 1D warp propagation theory coupled to a fast radiative transfer code that simulates the shadows induced. We consider two scenarios, namely a flyby in a plane orthogonal to, and at an angle with, the disc plane. In both models the outer disc becomes warped (leading to a broad shadow in the outer disc) and the warp wave propagates back and forth (causing the shadow to oscillate). We find that the inner disc, although tilted, is not warped and is therefore not shadowed. For a low viscosity disc ($\alpha=10^{-4}$) the warp lasts for most of the disc's lifetime ($\tau \sim 10^6\,$years), and for $50\%$ of the time the azimuthal variance of the surface brightness from the scattered light images, $\sigma^2$, is above $0.01$, meaning that the shadow in the disc is significant. We find that a significant fraction of discs in nearby star forming regions should have undergone a flyby sufficient to induce an observable warp, and that surveys of shadowed discs could provide a valuable probe of disc viscosity.
\end{abstract}


%%%%%%%%%%%%%%%%%%%%%%%%%%%%%%%%%%%%%%%%%%%%%%%%%%

%%%%%%%%%%%%%%%%% BODY OF PAPER %%%%%%%%%%%%%%%%%%

\section{Introduction}
\label{sec: intro}

Protoplanetary discs form as dense molecular clouds collapse to create stars (see \citealt{williams_2011} for a review). These discs are typically thin, planar structures composed of gas and dust orbiting the central star \citep{odell_1994}. In fact it was the coplanar nature of the solar system planets' orbits that led Kant and Laplace to conjecture that the solar system was formed from a disc \citep{kant_1755, de_laplace_1796}. However, the disc can be perturbed so that it is no longer planar and this has consequences for planet formation (such as planets forming with orbits misaligned with respect to the disc plane; \citealt{xiang_gruess_2016}). 

The most direct way to detect if a disc has perturbations in the disc plane is from shadows seen in scattered light images; stellar light scattering off small dust particles which are well coupled to the gas in the disc (although perturbations can be detected by other methods such as observations of disc kinematics eg. \citealt{winter_2025}). Scattered light images trace the surface of the disc so are sensitive to perturbations in the disc height which can block the stellar light from reaching outer parts of the disc, casting a shadow \citep[for a review see][]{benisty_2023}. A shadow is also cast if part of the disc is tilted so that the stellar light no longer impinges on the surface.

Recent observations have revealed that shadows are a common feature in scattered light images of protoplanetary discs \citep[eg.][]{stolker_2015, ginski_2021, debes_2023} suggesting that perturbations in protoplanetary discs are common. A misaglined disc, in which the inner disc is misaligned with respect to the outer disc, with a large inclination can cast two narrow shadows \citep[eg.][]{Marino_2015, Benisty_2017, Casassus_2018} as the stellar light can pass above and below the misaligned inner disc, but not through it. A misaligned disc with a lower inclination can cast a broad shadow across half of the outer disc \citep[eg.][]{Benisty_2018, Bohn_2019}. 

Broad shadows can be observed across more than half the outer disc \citep[][Milsom et al 2026 (in prep)]{muro_arena_2020} and this cannot be explained with just one misaligned annulus. Broad shadows can however be cast by a warp (where the disc plane is not constant with radius) in the disc. Examples of warped discs include Beta pic \citep{Burrows_1995,heap_2000}, AU Microscopii \citep{krist_2005}, and HD 100546 \citep{quillen_2006}. 

A warp propagates through the disc in a manner that depends the viscosity parameter, $\alpha$, \citep{shakura_sunyaev_1973} and the thickness of the disc, characterised by $H/R$. If $\alpha > H/R$ then viscous forces dominate and the warp propagation becomes diffusive \citep[the viscous regime, ][]{papaloizou_pringle_1983}. Alternatively, if $\alpha < H/R$ then pressure forces dominate and the warp propagates in a wave-like manner at half the sound speed of the disc ($c_{\rm s}/2$) \citep{papaloizou_lin_1995}. Current estimates of $\alpha$ for protoplanetary discs, $\alpha=10^{-4}$--$10^{-2}$ \citep{rosotti_2023}, suggest their warps should propagate in the wave-like regime.

If we treat the disc as a series of symmetric annuli then once an external torque has perturbed the disc then each annulus is not aligned with the next. This results in a pressure gradient between misaligned annuli (from the higher pressure region at the midplane of one annulus to the lower pressure region at the surface of the adjacent annulus) and the induced warp propagates through the disc \citep[see figure 10 from][and references therein]{Lodato_Pringle_2007}.

Warps can be induced due to an external torque perturbing the disc such as a misaligned magnetic field \citep{lai_1999}, a binary companion internal or external to the disc \citep{terquem_1993,paploizou_terquem_1995}, or by a flyby encounter: when an object on an unbound orbit passes nearby \citep[eg.][]{clarke_pringle_1993,Moeckel_2006,Nixon_Pringle_2010,xiang_gruess_2016,cuello_2019, kimmig_2026}. 

It is likely that a flyby encounter is a common occurrence in a disc's early lifetime as protoplanetary discs form around stars in chaotic and dense clusters \citep{lada_lada_2003, Porras_2003, Winter_2018, Galli_2019} and star-disc interactions are commonplace \citep{bate_2018}.  If an object passes sufficiently close to a disc it can tidally truncate the disc \citep{breslau_2014, bhandare_2016, breslau_2017}, or cause the disc to become highly inclined \citep{terquem_1993,terquem_1996,xiang_gruess_2016}, or warp the disc \citep{clarke_pringle_1993,Moeckel_2006,Nixon_Pringle_2010,xiang_gruess_2016,cuello_2019, kimmig_2026}.

The perturbation of the disc due to a flyby depends on whether its orbit is prograde/retrograde (the perturber orbits in the same/opposite direction as the disc rotates), or, in rare cases, orthogonal (the perturber orbit is perpendicular to the plane of the disc) \citep[see][for a review]{cuello_2023}. Perturbers with prograde orbits are more destructive than retrograde orbits due to resonance effects, and can cause tidal stripping and large spirals. Perturbers with retrograde orbits have less strong effects on the disc, but are more efficient at warping the disc and can cause misaligned discs. Once a warp has been induced by a perturbing flyby, it may last for a significant fraction of a disc's lifetime and we might expect to see evidence of this in observations. 

Shadows seen in scattered light observations of protoplanetary discs are often attributed to a misaligned or warped disc caused by a misaligned companion \citep[eg.][]{heap_2000,krist_2005, quillen_2006,Benisty_2018, Bohn_2019, nealon_2019,williams_2025}. Although there are studies that attribute disc structure to a recent or in fact ongoing flyby \citep[eg.][]{clarke_pringle_1993,Pfalzner_2003,Munoz_2014,cuello_2019}, until recently there have been few studies into the potential for a long-lasting, time-varying warps that can be caused by the flyby of a perturber much earlier in the disc's lifetime. However, \citet{kimmig_2026} explored the evolution of a flyby-induced warp demonstrating that although the induced spiral arms are short-lived the warp wave can persist for many thousands of years \citep[see also][]{Nixon_Pringle_2010}. These wave-like warps, which are a strong prediction of theory (see below), might be commonplace in discs and cause the misalignments seen in discs, or be an underlying process that is occurring alongside other  mechanisms causing large-scale scattered light structures. 

In this paper we explore the scattered light signatures of flyby induced warps by modelling the propagation of warps using 1D warp propagation theory and modelling scattered light images using a fast radiative transfer code. The magnitude of the warp depends on the mass, velocity, and closest approach of the perturber. We model the perturber with different closest approaches and velocities to test when the disc is significantly perturbed and what level of perturbation is observable. We also model the effects of a flyby encounter long after the perturber has passed to better understand the long-term impacts of flybys on protoplanetary discs.

The paper is structured as follows: in Section \ref{sec: modelling} we outline the 1D warp propagation theory behind our modelling and how we set-up our model. We also describe how we model scattered light images using a fast radiative-transfer code. In Section \ref{sec:results} we present the results of our models and in Section \ref{sec:discussion} we discuss the implications of our results for observations of warps induced by flybys. Finally, we compare our results with observations and discuss the likelihood and nature of flyby encounters.

\section{Modelling}
\label{sec: modelling}
In this section we introduce warp propagation theory, and the numerical methods we used to solve the wave equation. We also introduce our fast radiative transfer code, which we use to simulate the shadowing of warped discs.

\subsection{1D warp propagation theory}
\label{sec: model_theory}

We treat the disc as a series of annuli whose angular momenta are a function of radius only. This 1D approach enables us to rapidly compute models and explore parameter space, and to run models for long timescales. In contrast hydrodynamical models, which have been used successfully to model warped discs, are computationally expensive and are typically targeted at small numbers of initial conditions. 

For a small amplitude warp in the wave-like regime for a disc that is close to Keplerian (which is applicable to protoplanetary discs), the evolution of the warp can be described using the 1-D equations 12 and 13 from \citet{Lubow_Ogilvie_2000} \citep[or 4 and 5 from][]{Martin_2019}. These equations evolve the internal torque of the disc, $\textbf{\textit{G}}$, and the angular momentum vector, $\textbf{\textit{l}}$, of the disc through time, $t$:
\begin{align}
    \frac{\partial \textbf{\textit{G}}}{\partial t} - \omega\textbf{\textit{l}} \times\textbf{\textit{G}} + \alpha\Omega\textbf{\textit{G}} = \frac{\Sigma H^2R^3\Omega^3}{4}\frac{\partial \textbf{\textit{l}}}{\partial R}
	\label{eq: internal_torque}
\end{align}
and 
\begin{align}
    \Sigma R^2\Omega\frac{\partial \textbf{\textit{l}}}{\partial t} = \frac{1}{R}\frac{\partial \textbf{\textit{G}}}{\partial R} + \textbf{\textit{T}}
	\label{eq: angular_momentum}
\end{align}
where $\alpha$ is the Shakura-Sunyaev viscosity parameter, $\Omega$ is the angular frequency of the disc at a certain radius ($R$), $\Sigma$ is the surface density, $H$ is the height of the disc from the midplane, $\textbf{\textit{T}}$ is an external torque (for example due to a perturbing flyby), and $\omega$ is the apsidal precession frequency in the plane of the disc given by:
\begin{align}
    \omega = \frac{\Omega^2-\kappa^2}{2\Omega}
	\label{eq: apsidal_precession_frequency}
\end{align}
where $\kappa$ is the epicyclic frequency. We assume the disc to be in Keplerian rotation, for which $\Omega = \kappa$ and thus $\omega = 0$ in Equation \ref{eq: internal_torque}. The third term of Equation \ref{eq: angular_momentum} describes the damping of the warp due to viscosity as it propagates through the disc.

In this paper we focus on warps induced by a flyby encounter. For the external torque induced by flyby encounters we use the torque term from \citet{Lubow_Ogilvie_2000}:
\begin{align}
     \textbf{\textit{T}} = \frac{GM_{\rm p}\Sigma R}{2R^4_{\rm p}}\left[b^{(1)}_{3/2}\left(\frac{R}{R_{\rm p}}\right)\right](\textbf{\textit{R}}_{\rm p}\cdot\textbf{\textit{l}})(\textbf{\textit{R}}_{\rm p}\times\textbf{\textit{l}})
	\label{eq: flyby_torque}
\end{align}
where $G$ is the gravitational constant, $M_{\rm p}$ is the mass of the perturber, $\textbf{\textit{R}}_{\rm p}$ is the position vector of the perturber (at time $t$), and $b$ is the Laplace coefficient given by:
\begin{align}
     b_\gamma^{(m)}(x)= \frac{2}{\pi}\int^\pi_0 \cos(m\phi)(1+x^2-2x\cos\phi)^{-\gamma}d\phi
	\label{eq: laplace}
\end{align}

Equation \ref{eq: flyby_torque} describes the torque induced on the disc at each time-step due to a perturber at position $\textbf{\textit{R}}_{\rm p}$. Therefore to calculate the torque, we can calculate the path of the perturber (and so its position at each time-step). The direction of the torque is always perpendicular to the angular momentum vector of the disc and the position vector of the perturber.

From these equations we can model warps induced by a flyby encounter and how they propagate through the disc with time. We can see from Equations \ref{eq: internal_torque} and \ref{eq: angular_momentum} that $\textbf{\textit{G}}$ and $\textbf{\textit{l}}$ depend on each other so we must solve these equations simultaneously as outlined in the next section.

\subsection{1D warp model}
\label{sec: 1d_warp_model}

\begin{figure}
	\includegraphics[width=\columnwidth]{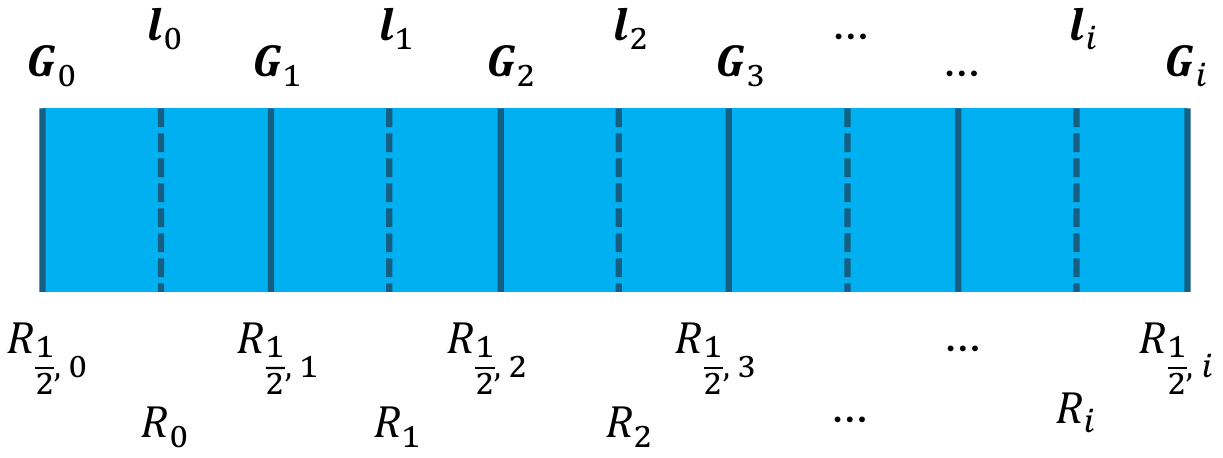}
    \caption{Figure showing the staggered grid set-up of our model. The blue represents the annuli which make up the disc where the solid lines show the edge of each annulus and the dotted lines show the centre of each annulus. $\textbf{\textit{l}}$ is defined at the full grid points ($R$) at the centre of each annulus and $\textbf{\textit{G}}$ is defined on the half grid points ($R_{\frac{1}{2}}$) at the edge of each annulus. The subscript denotes the position on the radial grid.} 
    \label{fig:staggered_grid}
\end{figure}

We calculate the angular momentum vector, $\textbf{\textit{l}}$, and internal torque,  $\textbf{\textit{G}}$, of the disc at each radius, for each time-step, using Equations \ref{eq: internal_torque} and \ref{eq: angular_momentum}. We set-up the disc so that it is in the $xy$-plane and calculate the tilt of the disc with respect to the $z$-axis from the angular momentum vector (as this is always perpendicular to the plane of the disc). Therefore we know the tilt of the disc at each radius and each time step and so how the warp propagates through the disc. 

We set up our disc using a staggered grid where $\textbf{\textit{l}}$ is defined on full grid points and $\textbf{\textit{G}}$ on half grid points (see Figure \ref{fig:staggered_grid}). This is because the angular momentum vector of the disc, $\textbf{\textit{l}}$, describes the tilt of each annulus and so is defined at the centre of each annulus (on the full grid points). In contrast, the internal torque of the disc, $\textbf{\textit{G}}$, describes the interaction between neighbouring annuli and is thus defined at the annulus boundaries (on the half grid points). We set the boundary conditions as $\textbf{\textit{G}}=0$ at $R_{\rm in}$ and $R_{\rm out}$ as there is no internal torque between annuli at the edge of the disc (where there is no adjacent annulus). 

We use Equations \ref{eq: internal_torque} and \ref{eq: angular_momentum} to model the disc, and Equation \ref{eq: flyby_torque} for the torque term (ignoring the $z$-component). We expect the warps to be relatively small so we can assume that $l_x,\ l_y\ll1$ and $l_z=1$ (since the perturbation from a flat disc, where $l_z=1$, is relatively small). Therefore we can focus on just the $x$- and $y$-components when modelling the disc.

To solve these equations we use the numerical method outlined in Section 4.1 of \citet{lubow_2002}.  We use a leap-frog method where $G1$ and $G2$, and $l1$ and $l2$ represent $G$ and $l$ respectively for different levels of the leapfrog (given by 1 and 2). To understand the methods used we can write Equations \ref{eq: internal_torque} and \ref{eq: angular_momentum} for the first level of the leap-frog (for the second level we simply swap all the 1s ands 2s) as:
\begin{align}
    \frac{G1^{j+1}-G1^j}{\Delta t} = -\alpha\Omega G1 + \frac{\Sigma H^2R_{(\frac{1}{2})}^3\Omega^3}{4}\frac{l2_i-l2_{i-1}}{{R_i}-R_{i-1}}
	\label{eq: internal_torque_method}
\end{align}
and
\begin{align}
    \Sigma R^2\Omega\frac{l1^{j+1}-l1^j}{\Delta t} = \frac{1}{R}\frac{G2_{i+1}-G2_i}{R_{_{(\frac{1}{2})i+1}}-R_{_{(\frac{1}{2})i}}} + T
	\label{eq: angular_momentum_method}
\end{align}
where $R_{(\frac{1}{2})}$ is the radial grid at half grid points ($R$ is at full grid points), $\Delta t$ is the difference between the current time step and the next time step, $j$ is an index for different time steps, and $i$ an index for different points on the radial grid. For example $G^j$ is the value of $G$ for the current time-step and $G^{j+1}$ is for the next time step. For values which $j$ and $i$ have been omitted they are the current time-step or radial point respectively (for clarity only the relevant values have been marked).

The first term of Equation \ref{eq: internal_torque_method} shows that we are calculating $G1$ for the next time step ($G1^{j+1}$) from the current time step ($G1^j$). The second term shows that viscosity term is updated from $G1$ using the forward Euler method and the third term that the radial derivative is updated from $l2$ using numerical differencing. Similarly, the first term of Equation \ref{eq: angular_momentum_method} shows that we are calculating the next time step for $l1$ (given by $l1^{j+1}$) from the current time step ($l1^j$). The second term shows that the radial derivative is updated from $G2$ using numerical differencing. We update the external torque term using the leap-frog method; so for the first level of the leap-frog (as in Equation \ref{eq: angular_momentum_method}) we update the torque from $l2$.

For Equation \ref{eq: internal_torque_method} the third term must be centred on a half grid point (even though $\textbf{\textit{l}}$ is defined on full grid points) as we are calculating $G$ (which is defined on half grid points). For example for $G1_{i=1}$, then we find the difference between $l2_{i=1}$ and $l2_{i=0}$ as this will be centred on the same point as $G1_{i=1}$ (see Figure \ref{fig:staggered_grid}). The same applies for the second term of Equation \ref{eq: angular_momentum_method}; it must be centred on full grid point as we are calculating $l$.

We validated our code by reproducing results from \citet{Martin_2019} (see their Figure 1) where they model the propagation of an arbitrary warp (ignoring the external torque term) in the wave-like regime (see Appendix \ref{sec:appendix_martin}). We also reproduced results from \citet{lubow_2002} (see their Figure 3) who modelled a disc perturbed by Lense-Thirring precession around a black hole (see Appendix \ref{sec:appendix_lubow}). Finally we reproduced models from \citet{Nixon_Pringle_2010} (see next section), who modelled how the disc is perturbed due to a flyby encounter (see Appendix \ref{sec:appendix_np10}). Satisfied that our model reproduced previous calculations, we can then proceeded to model the flybys.

\subsection{Flyby models}
\label{sec: flyby_models}
Using Equations \ref{eq: internal_torque} and \ref{eq: angular_momentum}, with Equation \ref{eq: flyby_torque} for the torque, we can model the propagation of warps induced by a flyby encounter. Since 1D warp modelling is fast it allows us to model the disc long after a flyby has passed; the main focus of this paper is the long-term effects of flyby encounters on protoplanetary discs rather than the intricate details of the flyby event itself, which can cause short lived spirals \citep{Pfalzner_2003, Munoz_2014, cuello_2019} or mass transfer \citep{clarke_pringle_1993,Munoz_2014, cuello_2019}. This is beyond the scope of this work.

\subsubsection{Model setup}
For our models we set-up the disc as outlined in section 3.3 of \citet{Nixon_Pringle_2010}. We assume the disc to have an outer radius of $R_{\rm out}=100\, \rm au$ and $R_{\rm in} = 1\, \rm au$. We note that \citet{Nixon_Pringle_2010} set the inner disc to $R_{\rm in} = 0.01\, \rm au$, however we reproduced their results for their model 1 with an inner disc of $R_{\rm in} = 1\, \rm au$ and found no significant difference between our models (see Appendix \ref{sec:appendix_np10}). We therefore adopt an inner radius of $R_{\rm in} = 1\, \rm au$ as this is less computationally expensive so allows us to run our models for longer and obtain results more quickly.

We run our models with 1000 radial grid points on a log-scale and set our time-step to satisfy the Courant-Friedrichs–Lewy condition \citep{courant_1928} given by:
\begin{align}
    C = u\frac{\Delta t}{\Delta R}
	\label{eq: courant_condition}
\end{align}
where $C$ is the Courant number, $u$ is the velocity of the perturber (as this is higher than the velocity of the warp), $\Delta t$ is the time step, and $\Delta R$ is the difference between radial grid points. $C$ must be equal to or smaller than 1, so we chose the minimum value of $\Delta t$ that satisfies this condition and divide it by 3 to ensure our time-steps are small enough.

We set-up the surface density to be a power law such that $\Sigma(R)\propto R^{-1}$. For our initial reproduction of model 1 from \citet{Nixon_Pringle_2010} (Appendix \ref{sec:appendix_np10}) we set the scale height to be $H/R=0.1$. However, we wish to visualise the shadows cast due to the warps induced (see Section \ref{sec:radiative_transfer_code}) so we model a flared disc in order that the shadows are visible in our scattered light models. Therefore we set up $H/R$ to have a canonical power law such that:
\begin{align}
    H = H_0\left( \frac{R}{R_0} \right)^\beta
	\label{eq: hr_power_law}
\end{align}t
where $H_0$ and $R_0$ are constants given by the height and radius at the outer disc respectively, and $\beta$ is the power law which describes how flared the disc is. We set $H_0=10\, \rm au$, at $R_0=100\, \rm au$, and $\beta$ to 1.125 (values that are typical for protoplanetary discs). We note that different scale heights affect how shadowed the disc is in scattered light and we explore this effect in Appendix \ref{appendix:h_effects}. We are in the wave-like regime so we set the disc to have a low viscosity with $\alpha = 10^{-4}$.

\subsubsection{Flyby setup}

\begin{figure}
	\includegraphics[width=\columnwidth]{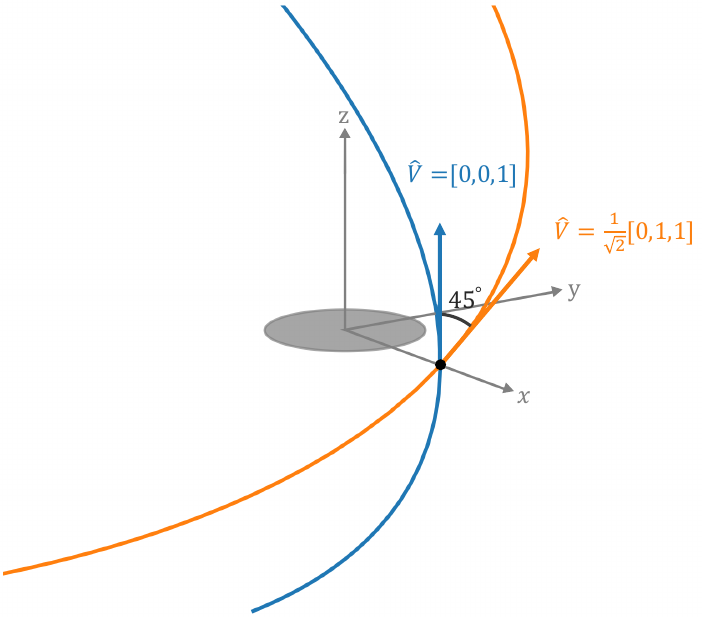}
    \caption{Figure showing the parabolic orbits of the perturbers for models 1 (blue line) and 2 (orange line). The disc is shown in grey (not to scale) in the $xy$-plane. The unit vector of the velocity is marked at periapsis for both orbits. Model 1 has a parabolic orbit in the $xz$-plane with periapsis at $[a,0,0]$ and $\mathbf{\hat{V}}=[0,0,1]$ at periapsis (so the perturber is travelling in the positive $z$-direction). Model 2 is the same shape orbit, but rotated $45^\circ$ clockwise about the $x$-axis so that the perturber is now travelling in a plane at $45^\circ$ the the $xz$-plane. The unit vector of the velocity at periapsis (which is still at $[a,0,0]$) is now $\mathbf{\hat{V}}=\frac{1}{\sqrt{2}}[0,1,1]$, rather than just in the $z$-direction.}
    \label{fig:hyperbolic_orbits}
\end{figure}

We setup our models of flyby encounters based on models 1 and 2 in \citet{Nixon_Pringle_2010} which model a perturber on a path perpendicular to the disc plane and at an angle to the disc plane. Models 1 and 2 from \citet{Nixon_Pringle_2010} assume that the perturbers follow a straight-line path, whereas we model the perturbers to be on a parabolic path (see Figure \ref{fig:hyperbolic_orbits}).

For our model 1 we set the perturber to be travelling on a parabolic orbit in the $xz$-plane travelling in the positive $z$-direction. The perturber passes through the disc plane ($xy$-plane) at its periapsis given by:
\begin{align}
    \textbf{\textit{R}}_{\rm peri} = [a, 0, 0]\, \rm au
	\label{eq: flyby_periaspsis}
\end{align}
where $a$ is the distance of the perturber from the disc centre at its closest approach. The perturber passes through the disc plane at $2000\,$years, with a velocity, $V$, in the positive $z$-direction ($\mathbf{\hat{V}}=[0,0,1]$) so at that moment it is travelling perpendicular to the disc plane. From these parameters, the whole parabolic orbit can be calculated. 

We set $a$ and $V$ to maximise the warp induced by the perturber; we set $a=300\,\rm au$ as this is the closest the perturber can be while still being above the disc truncation radius (below this radius the disc can become tidally truncated \citep{papaloizou_pringle_1977,paczynski_1977} and $V$ to be the escape velocity of the perturber at the closest approach (as this is the slowest the perturber can travel and escape the gravity of the host star):
\begin{align}
    V_{\rm esc} = \left(\frac{2G\mu}{a} \right)^{1/2} = 3.44\left(\frac{\mu}{2M_\odot} \right)^{1/2} \left(\frac{a}{300\, \rm au} \right)^{-1/2}\, \rm km\, s^{-1}
	\label{eq: flyby_velocity_model1}
\end{align}
where $\mu=M_{\rm s}+M_{\rm p}$ and $M_{\rm s}$ is the mass of the central star. We assume that the perturber has the same mass as the star ($M_{\rm p} = 1 M_\odot$). 

For our model 2 the perturber has the same shape path as for model 1, but it is now travelling at an angle to the $xz$-plane. We rotate the path for model 1 by $45^\circ$ clockwise about the $x$-axis so the perturber is travelling in a plane that is $45^\circ$ from the $xz$-plane (see Figure \ref{fig:hyperbolic_orbits}). The perturber still passes through the disc plane at $2000\,$years with $\textbf{\textit{R}}_{\rm peri} = [a, 0, 0]\, \rm au$ and $V=V_{\rm esc}$, but the direction of the velocity vector at periapsis is now given by $\mathbf{\hat{V}}=\frac{1}{\sqrt{2}}[0,1,1]$.

\subsection{Radiative transfer code}
\label{sec:radiative_transfer_code}

 In order to visualise the shadows in scattered light induced by disc warping we need to model the interaction of the stellar light with the dust in the disc. One possible method is to map the warped disc density structure onto a 3D grid and use an existing radiative transfer code to produce the synthetic observations. This is a potentially time-consuming step, with scattered light images using Monte Carlo methods requiring many minutes to run per timestep \citep[e.g.][]{Harries_2019, Sheehan_2017,williams_2025}.
 
Since we are  interested in just the scattered light we instead developed a fast, monochromatic radiative transfer code which models the scattered light only using a single-scattering approximation. The code does not perform a radiative equilibrium calculation and does not consider thermal emission. Although thermal emission from discs is important for the purposes of observational modes such as interferometry, here we are considering observations of large-scale scattering which for the most part will be obtained using adaptive optics observations combined with coronagraphy, which will obscure direct thermal radiation from the inner disc.

The first step is to map the warped disc density ($\rho$) structure onto a 3D spherical polar grid ($r, \theta, \phi$). The photospheric stellar radiation intensity at frequency $\nu$ is then propagated radially through the grid. We take $I_\nu = I_{*,\nu}$ (the emergent photospheric intensity) as the inner boundary condition, and numerically integrate the equation of radiative transfer, simplified by the above approximations:
\begin{equation}
\frac{dI_\nu(r,\theta, \phi))}{d\tau_\nu} = -I_{\nu}
\end{equation}
where $d\tau_\nu = (\kappa_{{\rm abs},\nu} + \kappa_{{\rm sca},\nu}) \rho(r,\theta,\phi) dr$, with $\kappa_{{\rm abs},\nu}$ and $\kappa_{{\rm sca},\nu}$ representing the monochromatic dust absorption and scattering opacities respectively. The mean intensity of the radiation field at each point is 
\begin{equation}
J_\nu(r,\theta,\phi) = I_{\nu} (r, \theta, \phi) W(r)
\end{equation}
where $W(r)$ is the geometric dilution factor from a star of radius $R_*$:
\begin{equation}
W(r) = \frac{R_*^2}{4 \pi r^2}
\end{equation}
which assumes $r \gg R_*$. Thus we obtain the mean intensity of the direct stellar radiation field for each voxel of the 3D grid.

The final part of the code takes the observer's position and direction ($\bf \hat{u}$) and creates an array of points in space which correspond to image pixels. The intensity to each of these pixels ($I_{{\rm obs},\nu}$) is calculated using a piecewise integration of the equation of radiative transfer through the grid towards the observer, starting with $I_{{\rm obs},\nu}=0$ as the ray enters the grid and solving:
\begin{equation}
\frac{d I_{{\rm obs},\nu}}{d\tau_\nu} = S_\nu - I_{{\rm obs},\nu}.
\end{equation}
Here $S_\nu$ is the scattering source function
\begin{equation}
S_\nu = \alpha_\nu J_\nu P(\mu, g)
\end{equation}
where $\alpha_\nu =\kappa_{{\rm sca},\nu} / (\kappa_{{\rm sca},\nu} + \kappa_{{\rm abs},\nu})$ is the albedo and $P(\mu, g)$ is the Henyey-Greenstein phase function \citep{Henyey_1941}:
\begin{equation}
P(\mu, g) = \frac{1}{4 \pi} \frac{ 1-g^2}{(1 + g^2 - 2g \mu)^{3/2}}.
\end{equation}
In the above $\mu$ is the cosine of the scattering angle, which in this case is $\mu = \bf{\hat{r}} \cdot \bf{\hat{u}}$,  and $g$ is the asymmetry parameter of the phase function such that $g = \langle \mu \rangle$.

For the models presented here the grid has 100 logarithmically-spaced radial points and 300 linearly-spaced azimuthal points. We  use 101 polar angle grid points, spaced such that
\begin{equation}
\theta_i = 
\begin{cases}
      \cos^{-1} \left( \left( \frac{i-51}{50} \right)^2 \right)  & \text{if $i \ge 51$},\\
       \pi -  \cos^{-1} \left( \left( \frac{i-51}{50} \right)^2 \right)  & \text{if $i<51$.}\\
    \end{cases} 
\end{equation}
We set the total disc mass (gas and dust) to be $0.01$\,M$_\odot$ and $R_* = 2$\,R$_\odot$. For our dust model we assumed silicate dust with an MRN size distribution \citep{Mathis_1977} and minimum and maximum grain sizes of 10\,nm and 1\,$\mu$m respectively, which is representative of the small grains dynamically coupled to the gas in the upper layers of protoplanetary discs. Assuming $J$-band observations we adopt a wavelength of 1.2\,$\mu$m. Using the optical properties from \cite{Draine_1984} our dust model yields a total dust opacity of $\kappa_{\rm tot}=78.5$\,cm$^2$g$^{-1}$ (per gram of gas, assuming a dust-to-gas ratio of 0.01) and an albedo of $0.84$. We assume isotropic scattering ($g=0$).

We tested the simplifying assumptions of our fast RT code by implementing our fiducial disc model in the Monte Carlo radiative transfer code TORUS \citep{Harries_2019}. The comparison between the fast and full RT codes is detailed in Appendix \ref{appendix:fast_rt_vs_full}. We show that the fast RT code compares very favourably with the results of the more detailed radiative transfer.

The code is written in Fortran 90 and parallelised using OpenMP, and is capable of producing a noise free $512 \times 512$ pixel image in a couple of seconds on a laptop. This  is approximately two orders of magnitude faster than a traditional Monte Carlo radiative transfer code, enabling us to examine the disc shadowing over a wide range of parameters, and for a large number of time-steps.

\section{Results}
\label{sec:results}

\begin{figure*}
	\includegraphics[width=\textwidth]{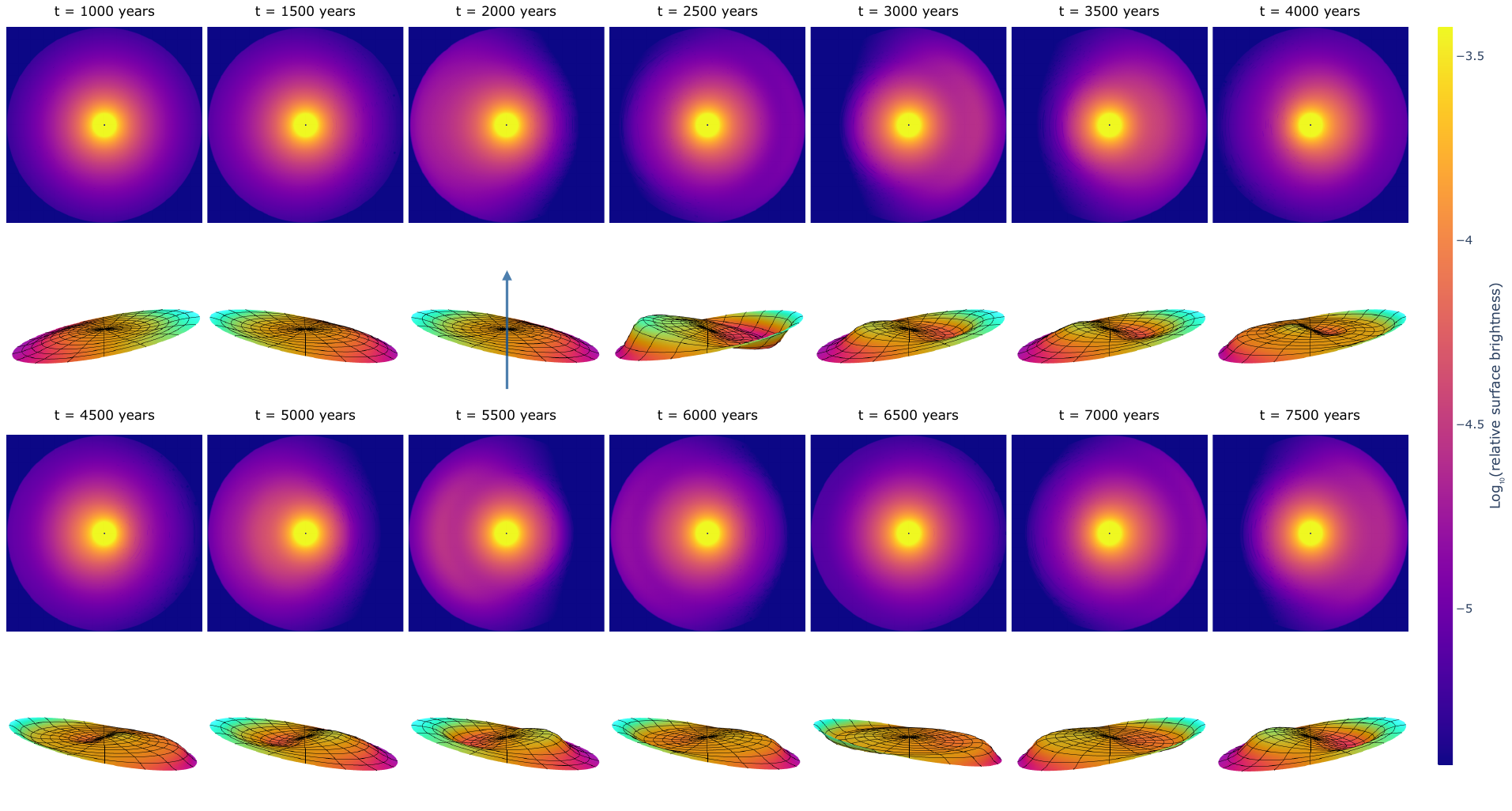}
    \caption{Shadows due to the warps induced by the flyby encounter from model 1. Rows 1 and 3 show face-on scattered light models produced using a fast radiative transfer code for different time periods. Each image is log scaled and normalised to the maximum surface brightness value. The image below each scattered light image shows the shaped of the warped disc where the colours show the height above (or below the midplane): orange is in the midplane, blue is above, and red below. These models are inclined by $70^\circ$ and the z-coordinates are exaggerated by a factor of 5, in order that the warp geometry is clear. The perturber passes the disc (its closest approach) at $2000\, \rm years$ (the direction of its path is show by the blue arrow) and here is where we see the largest tilt. The disc warps about the $x$-axis (as the flyby is travelling in the $xz$-plane) and the warp rocks back and forth about the $x$-axis. This produces a shadow which rocks back and forth.} 
    \label{fig:model_1}
\end{figure*}

\begin{figure*}
	\includegraphics[width=\textwidth]{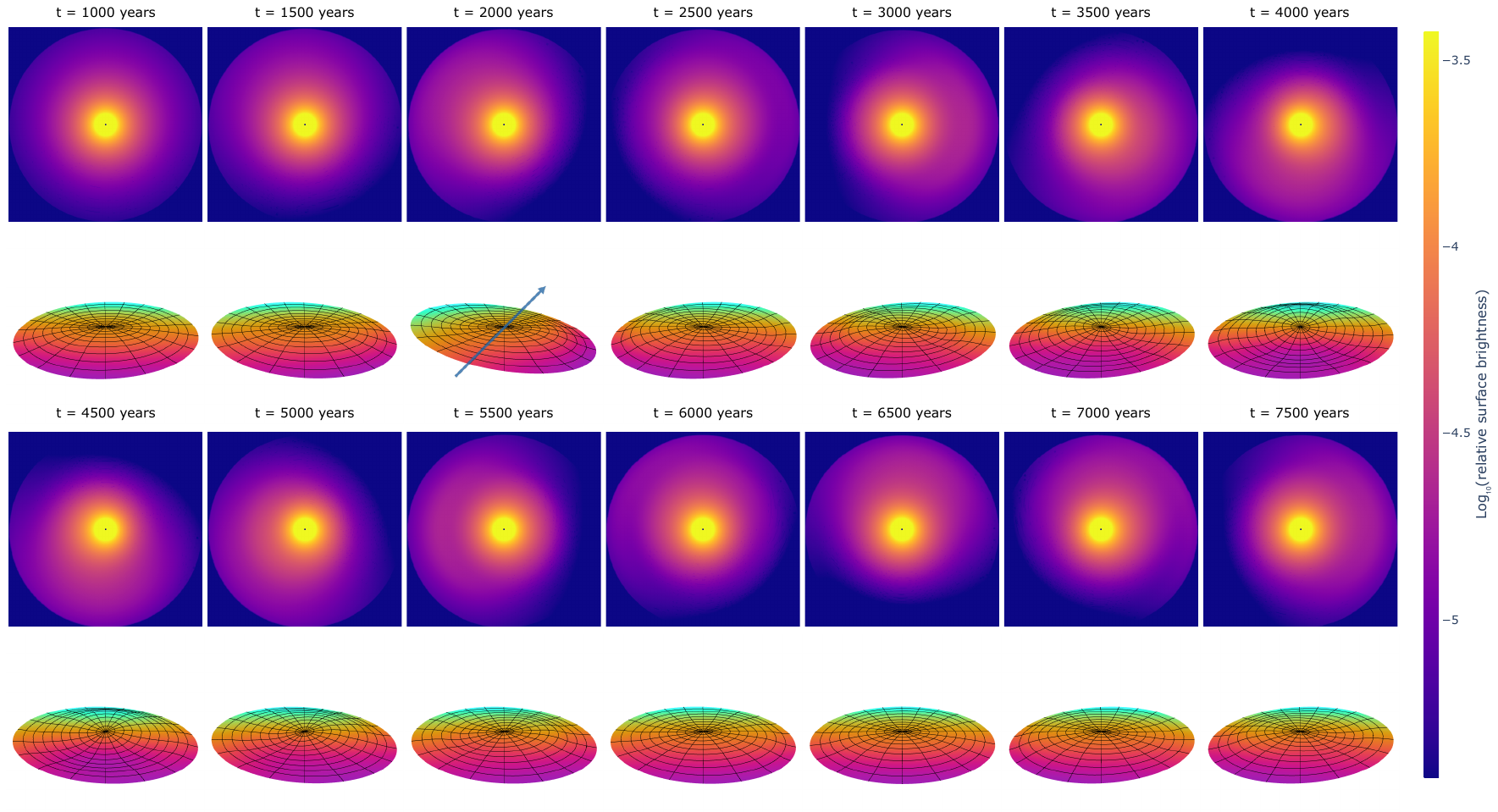}
    \caption{Same as figure \ref{fig:model_1}, but for the flyby encounter in model 2. Rows 1 and 3 show face-on scattered light models for different time steps (log scaled and normalised to the maximum surface brightness value) and the images below show the shaped of the warped disc where the colours show the height above (or below the midplane): orange is in the midplane, blue is above, and red below. These models are inclined by $70^\circ$ and the z-coordinates are exaggerated by a factor of 5, in order that the warp geometry is clear. The perturber passes the disc at its periapsis, at $\sim2000\,$years (the direction of the perturber is indicated by the blue arrow). Since the perturber is now travelling at an angle (rather than parallel with the $z$-axis) the induced warp now has a twist as well as a tilt. The shadow, due to the warp, moves anticlockwise round due to the twist before swapping sides of the disc and moving clockwise round once the perturber passes at $2000\,$years. Also, the whole disc tilts away from the $xy$-plane so that the North of the disc is always above the $xy$-plane, and the South below.} 
    \label{fig:model_2}
\end{figure*}

We present the results of model 1 in Figure \ref{fig:model_1}. The first and third rows show the face-on scattered light image produced from radiative transfer models of the disc for different time steps (normalised and log scaled). The rows below show the geometry of the disc where blue is above the $xy$-plane, red below, and orange in the $xy$-plane. These models are inclined by $70^\circ$ and the z-scale exaggerated by a factor of 5, so that the warp geometry is clear.

The perturber is initially in the negative $z$-direction and travels in the positive $z$-direction passing the disc plane ($xy$-plane) at 2000 years (the direction of the perturber is shown by the blue arrow in Figure \ref{fig:model_1}). The outer disc warps about the $x$-axis as the torque due to the perturber acts in a direction perpendicular to both $\textbf{\textit{l}}$ and $\textbf{\textit{R}}_{\rm p}$ (and the perturber is travelling in the $xz$-plane). As it passes the disc, the perturber reaches its closest approach ($x=a$ and $z=0$) and the disc warps in the opposite direction because the direction of the torque acting on the disc flips. The outer disc can be seen to warp back and forth about the $x$-axis even after the flyby has passed. A shadow is induced in the outer edge of the disc which is below the $xy$-plane. The side of the disc which is shadowed oscillates as the disc warps back and forth about the $x$-axis. The shadow is deepest when the tilt of the warped disc is largest.

It is worth noting that before the perturber passes the disc plane, the disc is already warped due to the perturber. The disc is warped at $1000\,$years, but then warps in the opposite directions at $1500\,$years. This is because the perturber has moved from having negative $x$-coordinates (at $1000\,$years) to positive (at $1500\,$years) so the direction of the torque acting on the disc changes. By $2000\,$years the warp is significant enough for a shadow to be induced in the outer edge of the disc which is below the $xy$-plane. 

The very inner disc ($R<35\,$au) does not warp, but tilts as a whole (relative to itself it stays flat). This is because the warp is induced in the outer disc due to the flyby and propagates inwards before reflecting back when it reaches the inner disc (causing the shadow to swap sides). The disc does not warp where the radius of the disc is less than the wavelength of the warp and the warp is reflected. Therefore the inner region of the disc does not become shadowed. This radius is given by $R_{\rm crit}\approx a(H/(2R))^{2/3}$ \citep{Nixon_Pringle_2010} and for $a=300\, \rm au$, $R_{\rm crit} \approx 41\, \rm au$. 

The results for model 2 are presented in Figure \ref{fig:model_2}. The perturber is initially in the negative $z$-direction and travels towards the disc perturbing it. The disc acquires a twist as the path of the perturber is no longer in the $xz$-plane, but is travelling at a $45^\circ$ angle to the disc plane. A shadow is induced in the warped outer disc which rotates anticlockwise (due to the twist) between $1500\,$ and $2000\,$years. Again, at $2000\,$years the perturber passes through the $xy$-plane (at $[a,0,0]$) and the direction of the tilt swaps. The shadow swaps to the other side of the disc and rotates clockwise (in the opposite direction to before). The twist is still present in the disc long after the perturber has passed and the shadow continues to rotate even after the perturber has passed the disc.

For model 1 the warp, and therefore shadow, oscillates back and forth about the $x$-axis, whereas for model 2 the shadow rotates. This is because for model 1 the torque mainly acts in the $y$-direction (see Equation \ref{eq: flyby_torque}) as the perturber is travelling in the $xz$-plane so the $\textbf{\textit{l}}$ vector oscillates back and forth about the $x$-axis. Whereas for model 2 the torque has both $x$- and $y$-components (as the perturber is travelling at an angle the the $xz$-plane). Therefore a twist is introduced as well as a tilt and the $\textbf{\textit{l}}$ is no longer limited to the $y$-direction. Also, the effects of the torque leads to a warp wave which means that $l_x$ and $l_y$ are $\sim90^\circ$ out of phase so the shadow rotates.

Additionally, for model 2 the whole disc tilts away from the $xy$-plane (see Section \ref{sec:asymmetry_parameter}, Figure \ref{fig:var_tilt_model1}) as shown by the second and third rows of Figure \ref{fig:model_2} which show the geometry of the disc. Even when the shadow is in the North of the disc (eg. at 4000 years) the North of the disc is always above the $xy$-plane, and the South below because the whole disc has tilted away from the $xy$-plane.

\subsection{Observability of shadows}
\label{sec:asymmetry_parameter}

\begin{figure*}
	\includegraphics[width=\textwidth]{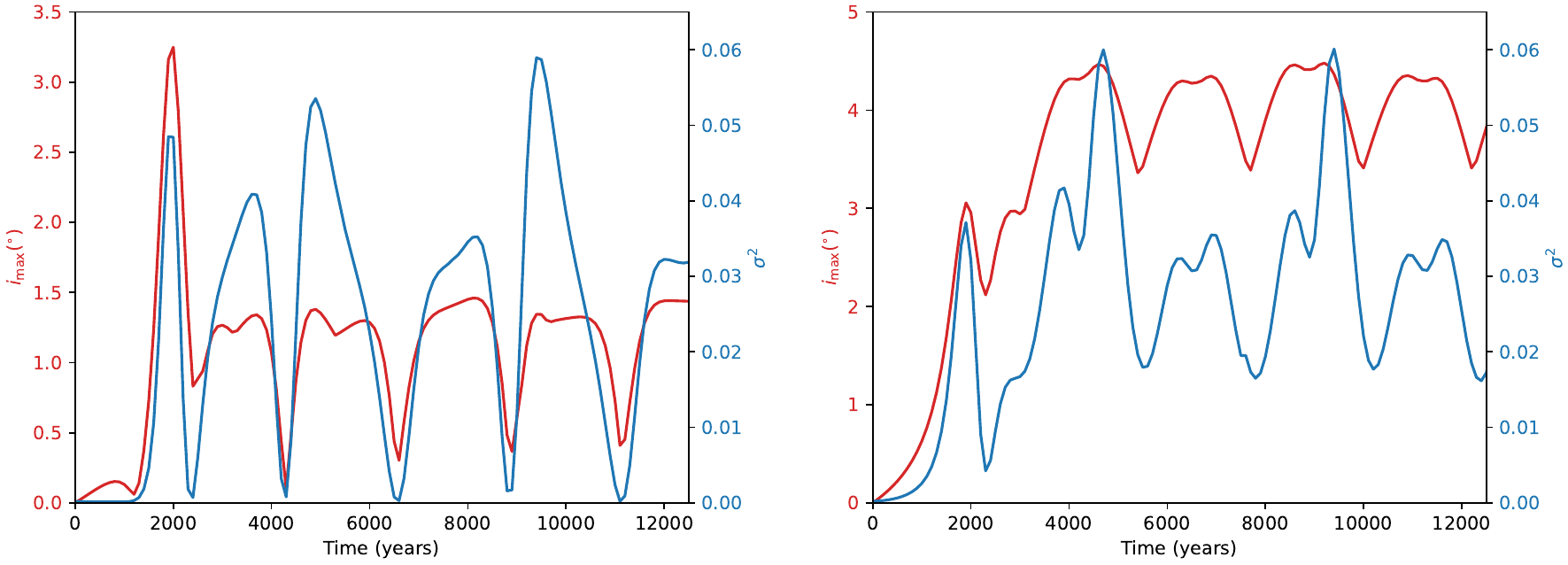}
    \caption{Figure showing the maximum tilt of the disc (red), and the azimuthal variance of the surface brightness of the scattered light image, $\sigma^2$, (blue) at each time step for models 1 (left) and 2 (right). The tilt is calculated as the angle between the angular momentum vector of the disc and the $z$-axis for each radius of the disc and the maximum value is taken for each time step. For model 1 the disc initially is warped at $2000\,$years as the perturber passes before settling into a an oscillating pattern of tilting back and forth. The lower the maximum tilt the less shadowed the disc and the lower the azimuthal variance. For model 2 we can see that the whole disc becomes tilted by the perturber so once the tilt settles into an oscillating pattern (after $\sim 3000\,$years) the minimum of $i_{\rm max}$ is not zero (i.e. the whole disc has tilted away from $xy$-plane). The shadow is a constant feature for model 2 as $\sigma^2$ does not drop to zero, unlike model 1 where $\sigma^2$ drops to zero (the disc is un-shadowed) as the warp oscillates back and forth.}
    \label{fig:var_tilt_model1}
\end{figure*}

In order to quantify the observability of the shadow induced in the disc we measure the azimuthal variance ($\sigma^2$) of surface brightness of the disc. This is a measure the spread of surface brightness values (azimuthally) from the mean surface brightness. An un-warped face-on disc should be azimuthally symmetric so have a low spread of surface brightness values and a low value of $\sigma^2$. However, if a shadow is induced in half the disc (as in model 1), there will be a large spread in surface brightness values azimuthally (at some angles the disc will be bright, but at others it will be shadowed) so $\sigma^2$ will be high. Therefore $\sigma^2$ is a good measure of how azimuthally symmetric the surface brightness of the disc is and so how observable shadows (which introduce an azimuthal asymmetry) may be. 

To measure the azimuthal variance, we split the disc into 30 azimuthal sections (between $R= 10\, \rm au$ and $R=R_{\rm out}$) and measure the mean surface brightness in each section. This gives us an azimuthal profile of mean surface brightness values (from each section) versus angle (given by the central angle of each section). We can then calculate the variance of the surface brightness using equation:
\begin{align}
    \sigma^2= \frac{\sum_i |x_i-\bar{x}|^2}{N}
	\label{eq: variance}
\end{align}
where $x_i$ is the mean surface brightness at a certain angle (from one azimuthal section), $\bar{x}$ is the mean surface brightness from all angles (from all azimuthal sections), and $N$ is the number of angular sections. 

Before calculating $\sigma^2$ we normalise the azimuthal profile to the mean surface brightness (from all angles). Since $\sigma^2$ measures the deviation from the mean, setting the mean to unity allows us to directly compare $\sigma^2$ across different images. This normalisation makes it straightforward to express $\sigma^2$ as a percentage of the mean. 

$\sigma^2$ is a good measure of how asymmetric the surface brightness of the disc is and can be compared for both observations and models. For the models in this paper the main feature seen in the scattered light models is a broad shadow in the outer disc; the images are otherwise fairly featureless. This allows us to be confident that for these models $\sigma^2$ is a measure of how shadowed the disc is. However for observations other features can contribute to an asymmetry in the disc (such as spirals) so care must be taken when using $\sigma^2$ to measure shadows in observations. The main purpose of this parameter is to give us a measure of how shadowed the disc is in our models due to warps. 

In Figure \ref{fig:var_tilt_model1} we see how $\sigma^2$ (blue) and maximum tilt of the disc, $i_{\rm max}$ (red), change with time for models 1 (left) and 2 (right). The tilt of the disc is calculated as the angle between the angular momentum vector of the disc and the $z$-axis. The tilt is calculated for each radius and $i_{\rm max}$ is the maximum of these tilts. For model 1, the greater $i_{\rm max}$, the greater $\sigma^2$ as the disc is more azimuthally asymmetric due to a larger shadow induced by the warp. There is a spike in $i_{\rm max}$ at $2000\,$years as this is where the perturber passes through the plane of the disc (its closest approach). After this the disc settles into a pattern of tilting back and forth and so $\sigma^2$ increases and decreases as the shadow moves across the disc.

For model 2 we see the disc is initially perturbed and then the tilt of the disc settles into an oscillating pattern. The whole disc becomes tilted by the perturber so the minimum of $i_{\rm max}$ after $\sim3000\,$years is not zero (as in model 1) as the whole disc tilts away from the $xy$-plane (see Figure \ref{fig:model_2}). Also, for model 1 the warp oscillates back and forth and the disc is relatively flat between oscillations ($i_{\rm max}$ drops to zero) as all the angular momentum vectors are pointing in the same direction at this point. This is not the case for model 2; since a twist has been induced as well as a tilt, the angular momentum vectors do not point in the same direction and there is no point where the disc is relatively flat. Therefore $\sigma^2$ does not drop to zero and a shadow is consistently present. This also contributes to why the tilt does not drop to zero.

We can use $\sigma^2$ to examine how varying parameters for our model affects the observability of shadows induced. We say that a shadow is observable if the standard deviation ($\sigma$) of the azimuthal surface brightness is more than $\sim 10 \%$ of the mean surface brightness (across all angles in the measured region). Below this value and the shadow would likely be indistinguishable from noise as estimated from scattered light observations of protoplanetary discs (\citealt{laws_2020, muro_arena_2020}; Milsom et al, in prep) from SPHERE \citep{beuzit_2019} and GPI \citep{macintosh_2006}. For example, we estimate the value of $\sigma$ to be $<10\%$ from noise for SPHERE images of the disc HD 139614 (whereas for the broad shadow in the outer disc $\sigma\sim 60\%$). Therefore we expect a shadow to be observable in a disc if $\sigma^2>0.01$ (see Appendix \ref{appendix:b} for comparisons of different $\sigma^2$ values).

\subsection{Varying \texorpdfstring{$V$}{V} and \texorpdfstring{$a$}{a}}
\label{sec:varying_v_a}

\begin{figure}
	\includegraphics[width=\columnwidth]{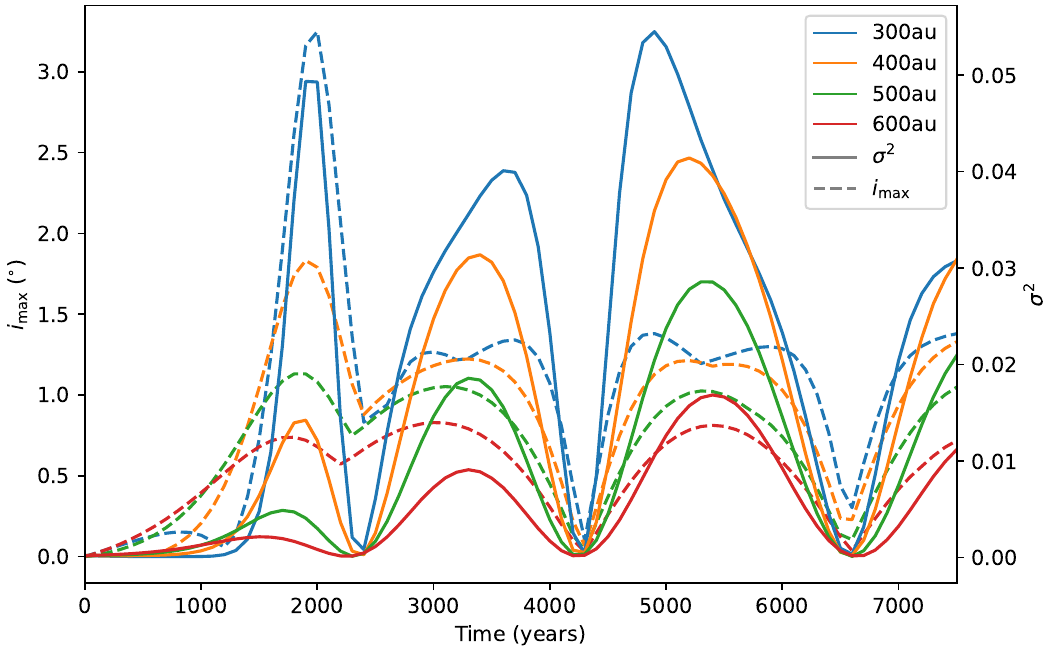}
    \caption{Same as the left of Figure \ref{fig:var_tilt_model1}, but for $a=300$, 400, 500, and $600\,\rm au$ (blue, orange, green, and red lines respectively) for model 1; $i_{\rm max}$ (dashed lines) and $\sigma^2$ (solid lines) are plotted against time. $V=V_{\rm esc}$ for each model. We can see that for lower values of $a$ the disc is less perturbed so less tilted and the scattered light images are less asymmetric (the shadows cast are less deep). This is because the torque on the disc is lower as the flyby is further away so the effects of its gravity are less.} 
    \label{fig:asymmetry_positions}
\end{figure}

\begin{figure}
	\includegraphics[width=\columnwidth]{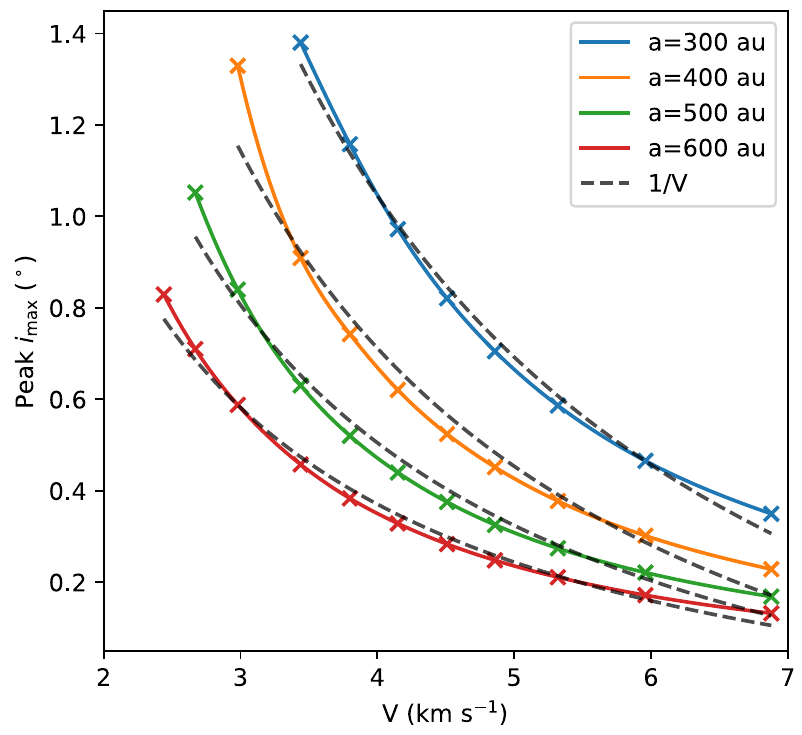}
    \caption{Figure showing the peak of the maximum tilt (after $2500\,$years), $i_{\rm max}$, for model 1 with varying values of $a$ and $V$. Models were run for 7500 years. $i_{\rm max}$ is plotted against $V$ for $a=300,\ 400,\ 500,$ and $600\,\rm au$ (blue, orange, green, and red lines respectively). Results from our models are plotted as crosses and the lines have been interpolated. We have also plotted lines proportional to $1/V$ (black dotted lines) to show that peak of $i_{\rm max}$ scales with $1/V$ as expected from our analysis of the angular impulse.}
    \label{fig:imax_vs_v_gridmodels}
\end{figure}

\begin{figure}
	\includegraphics[width=\columnwidth]{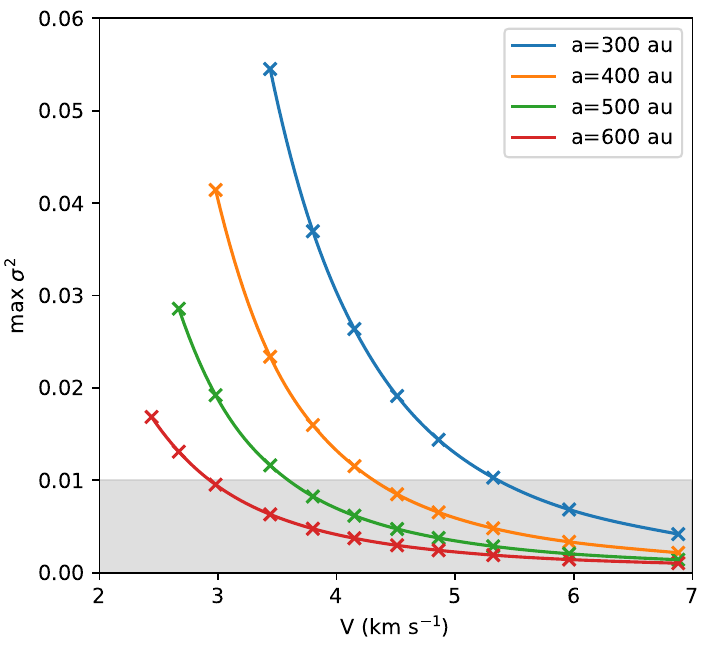}
    \caption{The peak of $\sigma^2$ (after $2500\,$years) for model 1 with different values of $V$ and $a$ of the perturber; $a=300,\ 400,\ 500,$ and $600\,\rm au$ (blue, orange, green, and red lines respectively. Results from our models are plotted as crosses and the lines have been interpolated. The figure shows the same models as in Figure \ref{fig:imax_vs_v_gridmodels}). The grey region marks where $\sigma^2<0.01$ and shadows are no longer observable.}
    \label{fig:a_v_tilt_model1}
\end{figure}

\begin{figure}
	\includegraphics[width=\columnwidth]{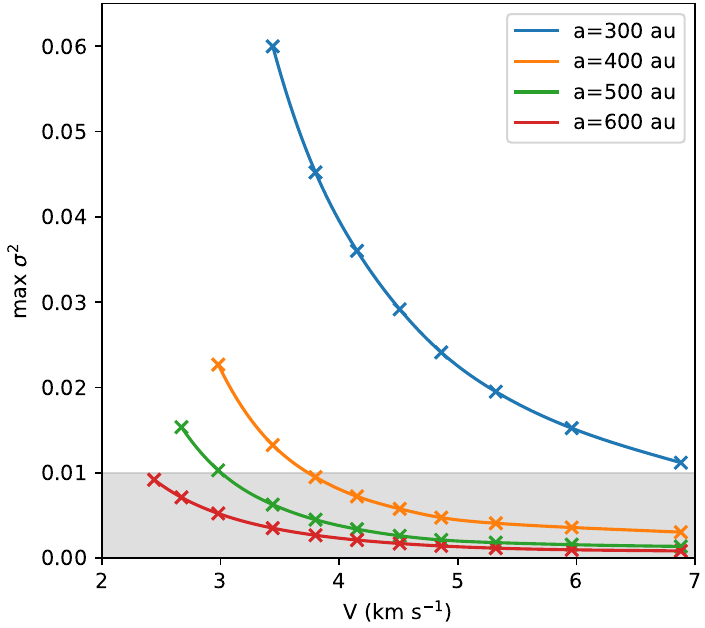}
    \caption{Same as Figure \ref{fig:a_v_tilt_model1}, but for model 2. The peak of $\sigma^2$ is plotted against $V$ for $a=300,\ 400,\ 500,$ and $600\,\rm au$ (blue, orange, green, and red lines respectively). The grey region marks where $\sigma^2<0.01$ and shadows are not observable.}
    \label{fig:a_v_tilt_model2}
\end{figure}

In this section we explore how the warp induced in the disc and therefore the shadows induced depend on the velocity and closest approach of the perturber. We vary these parameters for the path of the perturber for both models 1 and 2.

To determine how changing $V$ and $a$ affect how warped the disc becomes we can calculate the angular impulse, $\textbf{\textit{L}}$, on the disc due to the perturber:
\begin{align}
    \textbf{\textit{L}} = \int\textbf{\textit{T}}dt
	\label{eq: angular_impulse}
\end{align}
This quantifies the change in angular momentum of the disc, providing a measure of how much torque has acted on the disc and hence how much the disc has been perturbed. 

To calculate the angular impulse, we can use a simplified version of the torque in Equation \ref{eq: flyby_torque} from \citet{Nixon_Pringle_2010}\footnote{Note a factor of two is included in the denominator which was missed in the original derivation from \citet{Nixon_Pringle_2010}.}:
\begin{align}
     \textbf{\textit{T}} = \frac{3GM_{\rm p}\Sigma R^2}{2(R^2_p + R^2)^{5/2}}(\textbf{\textit{R}}_{\rm p}\cdot\textbf{\textit{l}})(\textbf{\textit{R}}_{\rm p}\times\textbf{\textit{l}})
	\label{eq: flyby_torque_nixon}
\end{align}
which uses an approximation for the Laplace coefficient. For model 1, using our assumption that $l_x,\ l_y\ll1 $ so $l\sim[0,0,1]$, and integrating between $t=t_0$ and $t=\infty$ we calculate the magnitude of $\textbf{\textit{L}}$ to be:
\begin{align}
    L = \frac{GM_{\rm p}\Sigma R^2 a}{2V(a^2+R^2)^{\frac{3}{2}}}\sim \frac{GM_{\rm p}\Sigma R^2}{2Va^2}
	\label{eq: angular_impulse_model1}
\end{align}

From this equation, if $a$ is constant, $L\propto\frac{1}{V}$. Hence if the velocity of the perturber is increased, the angular impulse on the disc is decreased. This is because the perturber spends less time in close proximity to the disc perturbing it, so overall the torque induced on the disc is lower. Similarly, if $V$ is held constant, $L\propto\frac{1}{a^2}$; increasing $a$ decreases the angular impulse on the disc and the disc becomes less warped. 

We can see the maximum tilt of the disc decrease as $a$ decreases in Figure \ref{fig:asymmetry_positions} which shows how $i_{\rm max}$ and $\sigma^2$ vary with time for $a=300$, 400, 500, and $600\,\rm au$ (blue, orange, green, and red lines respectively) for model 1 (with $V=V_{\rm esc}$). $a=300\, \rm au$ is the minimum value of $a$ that we model as this is at the truncation radius of the disc. If we set $a=600\, \rm au$ (double the truncation radius) then the peak of $i_{\rm max}$ of the disc (at $\sim3000\,$years) is $\sim2$ times lower than for $a=300$ and similarly the peak of $\sigma^2$ is $\sim4$ times lower. The maximum tilt of the disc, and so azimuthal asymmetry due to shadowing, is much lower.

Figure \ref{fig:imax_vs_v_gridmodels} shows how varying $V$ and $a$ affects how warped the disc becomes for model 1. The peak of $i_{\rm max}$, after $2500\,$ years (after the initial perturbation of the disc), is plotted against $V$ for $a=300,\ 400,\ 500,$ and $600\, \rm au$ (blue, orange, green, and red lines respectively). Values for $V$ were chosen to be $V_{\rm esc}$ at $a=300$, 400, 500, $600\,$au $\sim3.44$, 2.98, 2.67, $2.44\, \rm km\, s^{-1}$; twice these values; and values evenly sampled in between $V_{\rm esc}(a=300\, \rm au)$ and $2V_{\rm esc}(a=600\, \rm au)$. Each model was run for $7500\,$years. We have also plotted lines proportional to $1/V$ (black dotted lines), confirming that the models are proportional to $1/V$ as expected from the calculations above. From these plots we see that the greater $a$ and $V$ the lower the peak of $i_{\rm max}$ and so the less warped (and therefore less shadowed) the disc becomes. The disc is most warped at the lowest values of $V$ and $a$. 

Also, for a constant value of $V$, decreasing $a$ increases $i_{\rm max}$. However it is worth pointing out that decreasing $a$ also acts to increase $i_{\rm max}$ even when $V=V_{\rm esc}(a)$ (we can see this by looking at the models with the lowest values of $V$ for each coloured line or from Figure \ref{fig:asymmetry_positions}). Decreasing $a$ and $V$ acts to increase $L$. However, by using $V_{\rm esc}$ we are increasing $V$ as we decrease $a$ (see Equation \ref{eq: flyby_velocity_model1}) which acts to decrease $L$. Overall though $L$ has a stronger dependence on $a$ than $V$ and for $V=V_{\rm esc}$, $L\propto\frac{1}{a^{3/2}}$ so decreasing $a$ still increases $L$.

Figures \ref{fig:a_v_tilt_model1} and \ref{fig:a_v_tilt_model2} show how the peak of $\sigma^2$ (after $2500\,$years) varies when changing $a$ and $V$ for models 1 and 2 respectively. $\sigma^2$ quantifies how shadowed the disc is (see Section \ref{sec:asymmetry_parameter}), so these plots show how shadowed the disc becomes for different values of $a$ and $V$. The peak of $\sigma^2$ is plotted against $V$ for $a=300$, 400, 500, and 600 (blue, orange, green, and red lines respectively). The grey region marks where $\sigma^2<0.01$ and so the shadows are not observable (see Section \ref{sec:asymmetry_parameter}). We note that Figure \ref{fig:a_v_tilt_model1} shows the same models as those in Figure \ref{fig:imax_vs_v_gridmodels}.

From these plots we conclude that the disc is most shadowed at the lowest values of $V$ and $a$ as this is when the disc is most warped (see Figure \ref{fig:imax_vs_v_gridmodels}). $\sigma^2$ follows the same trends as $i_{\rm max}$: it is highest when $V$ and $a$ are lowest, and increasing $a$ increases $\sigma^2$ for constant $V$ and $V=V_{\rm esc}$. This is because the disc is most shadowed (as measured by $\sigma^2$) when it is most warped (as measured by $i_{\rm max}$).

For model 1 the warp oscillates back and forth about the $x$-axis even for different values of $V$ and $a$. However, for model 2 the shadow rotates when $a=300\,$au, but not for larger values of $a$ (instead shadow oscillates in a manner more similar to model 1). This is because whether the shadow rotates depends on whether $l_x$ and $l_y$ are $\sim90^\circ$ out of phase (see Section \ref{sec:results}) which is not always the case. We can see from Figure \ref{fig:a_v_tilt_model2} that when $a=300\,$au the maximum of $\sigma^2$ is high and a large shadow is induced in the disc, but there is a sudden drop in the maximum of $\sigma^2$ when $a>300\,$au and a smaller shadow is induced which does not rotate.

Varying these parameters allows us to see how they affect the magnitude of the warps induced, and so the depth and observability of the shadow in the scattered light images. To understand the observability of warps induced by flybys more fully we need to understand the timescale of the warps and whether we are likely to observe such a warped disc.

\subsection{Timescale of warping}
\label{sec:timescale_of_warps}
\begin{figure}
	\includegraphics[width=\columnwidth]{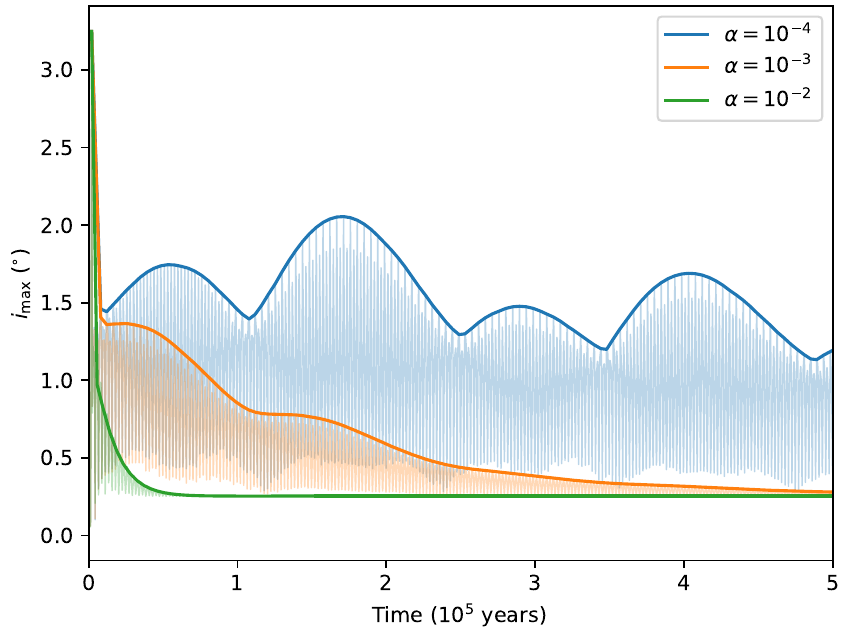}
    \caption{Figure showing the peak of the maximum tilt for each oscillation of the warp with time. Oscillations of the maximum tilt of the disc with time are plotted (low opacity lines) for model 1 with $\alpha=10^{-4},\ 10^{-3},\ 10^{-2}$ (blue, orange, and green lines respectively). The low opacity blue line is the same as the red line in the left of Figure \ref{fig:var_tilt_model1} but for $10^5\,$years. The peak of each oscillation is then plotted as a solid line. The more viscous the disc (the higher the value of $\alpha$) the quicker the warp of the disc decays. However for $\alpha=10^{-4}$ (blue line) the disc warping occurs on a much longer timescale. We ran this simulation for $10^6\,$years and the disc was still warped with a maximum tilt above $0.9^\circ$.}
    \label{fig:viscosity_timescales}
\end{figure}

\begin{figure}
	\includegraphics[width=\columnwidth]{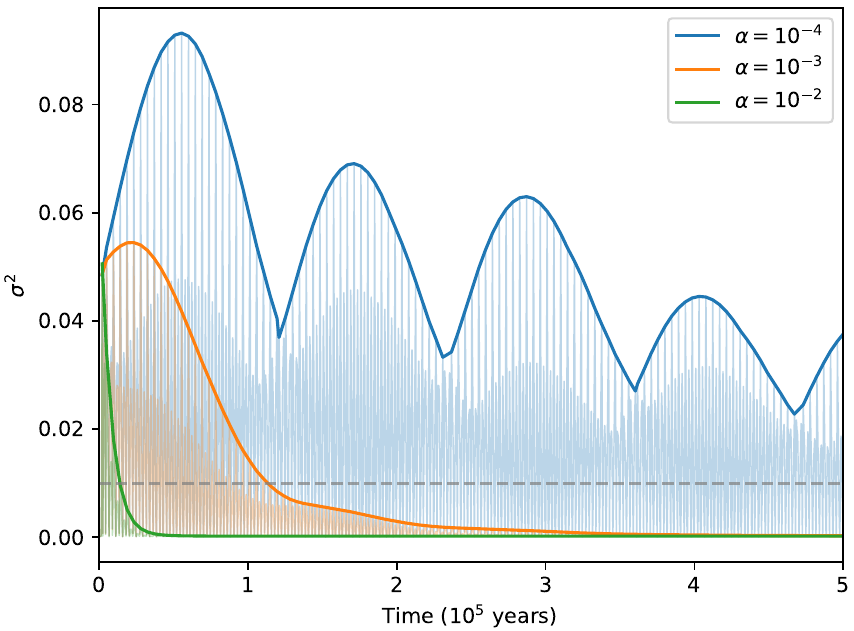}
    \caption{Figure showing $\sigma^2$ (low opacity lines) with time for model 1 with $\alpha=10^{-4},\ 10^{-3}$, and $10^{-2}$ (blue, orange, and green lines respectively). The peak of $\sigma^2$ for each oscillation is plotted as solid lines. The grey dotted line marks where $\sigma^2=0.01$ where below this line shadows are no longer observable. The higher the viscosity of the disc the quicker the warp decays and the shadows become unobservable. For a low viscosity disc ($\alpha=10^{-4}$) we ran this model for $10^6$ years and the shadows were still observable ($\sigma^2>0.01$) even after $10^6$ years.} 
    \label{fig:sigma2_timescales}
\end{figure}

 Determining the lifetime of the induced warp can be used to estimate how long we may expect the disc to be shadowed and therefore how likely it is that we may observe shadowing due to a flyby. Also, measuring the period of oscillation of the warp allows us to determine the speed of the warp propagation whether we could detect temporal variation in the shadows between epochs of observations.

\citet{lubow_2002, Nixon_Pringle_2010} approximate the decay timescale for a warp as:
\begin{align}
    t_{\rm damp} \approx \frac{P_{\rm out}}{2\pi\alpha}
	\label{eq: warp_lifetime}
\end{align}
where $P_{\rm out}$ is the period of rotation at the outer edge of the disc.

In Figure \ref{fig:viscosity_timescales} we show $i_{\rm max}$ at each time step (low opacity lines) for $\alpha=10^{-4}$, $10^{-3}$, and $10^{-2}$ (blue, orange, and green lines respectively) for model 1. The maximum tilt of the disc for that oscillation, is plotted as a solid line. We fitted an exponential to this solid line and determined the mean lifetime $\tau$ (the time for the peak tilt to reduce by $1/e\approx0.37$). From Equation \ref{eq: warp_lifetime} for $\alpha=10^{-4}$, $t_{\rm damp}\sim 1.6\times10^{6}\,$years which matches our value of $\tau\sim10^6$ (we also obtained this value for model 2). Also, from our results the lifetime of the warp decreases by a power of 10 as $\alpha$ increases by a power of 10 which follows the relationship that $t_{\rm warp} \propto 1/\alpha$ as defined above. 

The greater the viscosity of the disc the more the warp is damped and so the shorter its lifetime. For $\alpha=10^{-2}$ the warp decays completely within $\sim 10^5\,$years and $\tau\sim10^4\,$years; the lifetime of the warp is very short and so it is unlikely that a warp due to a flyby would be observed for a disc with this viscosity (see Section \ref{sec:discussion} for a discussion on the likelihood of a flyby encounter and observing a warp due to a flyby). If a warped disc with this viscosity were to be observed then we would expect the flyby event to have happened recently and the perturber should be easily identifiable. For $\alpha=10^{-3}$ the warp decays within $\sim5\times10^5\,$years and $\tau\sim10^5\,$years. However, for the lowest viscosity disc with $\alpha = 10^{-4}$ we ran our model for $10^6\,$years and the disc was still warped with the peak of the maximum tilt above $0.9^\circ$. This means that for the lowest viscosities the warp would be present for a large part of the disc's lifetime (the median disc lifetime is 2-3 Myr: see \citealt{williams_2011} for a review). 

Figure \ref{fig:sigma2_timescales} shows the azimuthal variance of the scattered light images, $\sigma^2$, with time for model 1 (same as Figure \ref{fig:viscosity_timescales}, but for $\sigma^2$ rather than $i_{\rm max}$). $\sigma^2$ tells us how visible the shadows are in the disc (whereas $i_{\rm max}$ tells us how warped the disc is). $\sigma^2$ is plotted against time for model 1 with $\alpha=10^{-4}$, $10^{-3}$, and $10^{-2}$ (see legend) as a low opacity line. The peak of $\sigma^2$ for each oscillation is plotted as a solid line. From this plot we can see that the peak of $\sigma^2$ decreases more quickly for high viscosity discs as the warp is damped more quickly. These results match that as for $i_{\rm max}$ because the less tilted (or warped) the disc, the less shadowed. However, the mean lifetime of the peak of $\sigma^2$ (the time for the peak of $\sigma^2$ to reduce by $1/e\approx0.37$), $\tau_\sigma$, is about a magnitude lower than for $i_{\rm max}$ ($\tau_\sigma\sim 10^{5},\ 10^{4},\ 10^{3}\,$years for $\alpha=10^{-4},\ 10^{-3},\ 10^{-2}$ respectively). We also calculated $\tau_\sigma=10^5\,$years for model 2 with $\alpha=10^{-4}$.

\subsubsection{Timescale of oscillations}

\begin{figure}
	\includegraphics[width=\columnwidth]{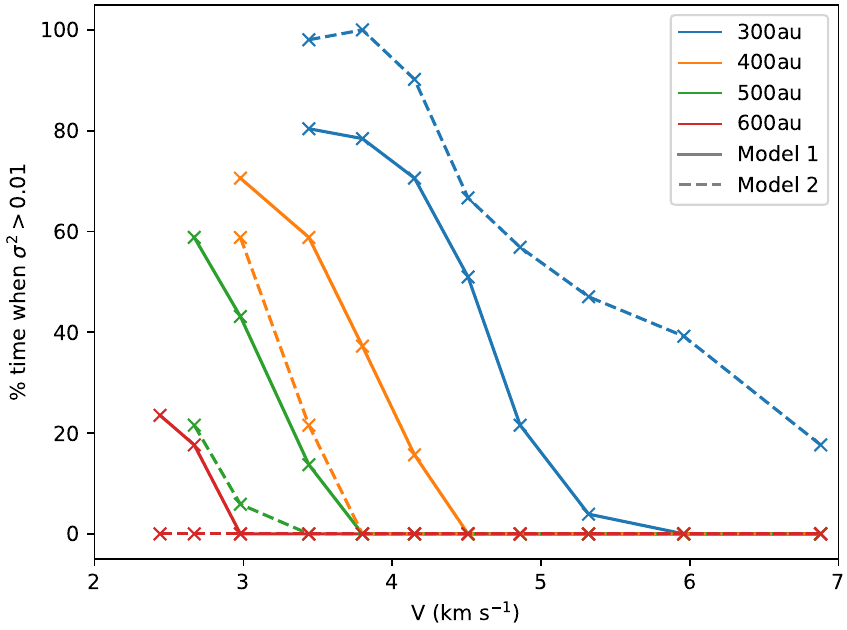}
    \caption{Figure showing the percentage of time when $\sigma^2>0.01$ (after $2500\,$years), and so shadows are observable, for models with varying $a$ and $V$ (see legend). All models were run for $7500\,$years so we are calculating the percentage of time which the disc is shadowed for $\sim1$ oscillation of the warp. Solid lines show results from the models based on model 1 which are shown in \ref{fig:a_v_tilt_model1}. Dashed lines show results from models based on model 2 which are shown in Figure \ref{fig:a_v_tilt_model2}.}
    \label{fig:sigma_greater_01}
\end{figure}

We have determined that for a disc with a low viscosity, warping due to a flyby encounter can last a similar time to the lifetime of the disc. However since the warp is oscillating (from tilted to flat to tilted the other way and back), for a large fraction of time it may not be warped sufficiently to produce an observable effect in scattered light. Measuring how much time that the disc is warped significantly allows us to calculate how likely we are to observe a disc perturbed by a flyby encounter at a moment in time where the perturbations are observable.

The period of one oscillation of the warp is given by:
\begin{align}
    P=4\int_{0}^{R_{\rm out}}\frac{dR}{c_s(R)}=\frac{32}{11}\frac{R_{\rm out}}{c_s(R_{\rm out})}
	\label{eq: oscillation_period}
\end{align}
\citep[based on equation 28 from][]{Nixon_Pringle_2010} using $c_s=H\Omega$ and $H$ from Equation \ref{eq: hr_power_law}. The factor of 4 comes from the fact that during one oscillation the warp travels twice $R_{\rm out}$ and travels at a speed of $c_s/2$. For model 1 this equation gives $P\approx4600\,$years. We measured the time period for an oscillation from our results by taking a Fourier transform of $i_{\rm max}$ (Figure \ref{fig:viscosity_timescales}), finding the average period for one oscillation of the disc to be $4546\pm1\,$years which is in agreement with our estimate above.

During the first oscillation of the warp for model 1, the disc spends $\sim800\,$years where $\sigma^2<0.01$ (from Figure \ref{fig:sigma2_timescales}) and so shadows are not visible. This occurs in the two troughs between the alternate peaks where the disc flattens as the direction of the tilt swaps. Thus for the first oscillation, the disc spends $\sim20\%$ of the time with $\sigma^2<0.01$, but $\sim80\%$ of the time the warp is significant enough for shadows to be observed.

As the maximum tilt of the warp decreases over time, the percentage of time that $\sigma^2>0.01$ decreases; after one mean lifetime of the warp ($\tau\sim 10^6\,$years) the maximum tilt has decreased from $\sim2^\circ$ to $\sim1^\circ$ and $\sigma^2>0.01$ has decreased from $\sim80\%$ of the time to $\sim30\%$. Overall, over one lifetime of the warp, the disc is significantly warped so that the shadows are visible ($\sigma^2>0.01$) for $\sim50\%$ of the time.

This behaviour also holds for a disc with higher viscosity ($\alpha=10^{-3}$). The time period of the oscillations remains the same (it does not depend on $\alpha$) so initially the time spent in the troughs between peaks is the same and $\sim80\%$ of the time the warp is significant and shadows are visible. Over one mean lifetime ($\tau\sim10^5\,$years), the disc is significantly shadowed, with $\sigma^2>0.01$, $\sim50\%$ of the time. For $\alpha=10^{-2}$ the warp is damped quickly so although within one mean lifetime ($\tau\sim10^4\,$years) $\sigma^2>0.01$ for $\sim70\%$ of the time, the warp decays rapidly and after one mean lifetime $\sigma^2$ is never above $0.01$ as the disc is not warped enough and so shadows are not visible.

We also calculated that for model 2 with $\alpha=10^{-4}$ for $\sim 80\%$ of the time over one lifetime of the disc $\sigma^2>0.01$ (compared to $50\%$ for model 1). This is because the shadow rotates and is constantly present until the warp decays so it is not observable. Unlike for model 1 where the shadow oscillates back and forth and is unobservable between oscillations.

In Figure \ref{fig:sigma_greater_01} we present the percentage of time when $\sigma^2>0.01$, and so the disc is significantly shadowed, for different values of $a$ and $V$ for models 1 (solid lines) and 2 (dashed lines). These are the same models as presented in Figures \ref{fig:a_v_tilt_model1} and \ref{fig:a_v_tilt_model2} (see Section \ref{sec:varying_v_a}). The percentage time was calculated after the initial tilt of the disc ($2500\,$years) and the models were run for $7500\,$years so these results show how shadowed the disc is initially. We focus on how shadowed the disc is on a shorter timescale of $\sim1$ oscillation. As discussed in Section \ref{sec:timescale_of_warps}, over longer time periods the warp is damped so this percentage would decrease for later oscillations (although this occurs slowly for low $\alpha$).

As discussed above, for model 1 the shadow oscillates back and forth and so $\sigma^2>0.01$ for less than $100\%$ of the time as between oscillations the disc is un-shadowed. However, for model 2 with $a=300\,$au and low values of $V$ the shadow rotates and is always present therefore $\sigma^2>0.01$ for $\sim100\%$ of the time (whereas $\sigma^2$>0.01 for $\sim80\%$ of the time for corresponding discs for model 1). For higher values of $V$ (with $a=300\,$au) although the shadow still rotates, it is weaker and not always observable. For $a>300\,$au for model 2 the shadow no longer rotates, but oscillates (see Section \ref{sec:varying_v_a}) so is not always present and $\sigma^2>0.01$ for a much less time.

It is worth noting that we observe longer time scale (lower frequency) oscillations of the warp for $\alpha=10^{-4}$; the solid blue line in Figure \ref{fig:viscosity_timescales} oscillates (as well as the transparent blue lines). From our analysis of our Fourier transformed data, we found that a secondary warp with a frequency very close to the primary warp is also induced and that these warps constructively interfere at the time period of the longer oscillations (the beat frequency of the two waves is the frequency of the longer oscillations). This secondary warp is visible as a second peak where the two warps are most out of phase. It can be seen as a double peak in Figure \ref{fig:var_tilt_model1}.

We note that $i_{\rm max}$ does not return to $0^\circ$ once the warp has dissipated (this is especially clear for the high viscosity case - green line in Figure \ref{fig:viscosity_timescales}). We attribute this to the fact that the disc is warped when the perturber passes so the torque acting on the disc is not symmetric (i.e. as the perturber passes the disc its path is not orthogonal to the disc plane since the disc has already been warped). Another, smaller, contributing factor is that we start the perturber at a finite distance from the disc, but once it passes we allow it to travel a large distance so the cumulative torque on the disc is not exactly balanced.

\subsection{Shadowing at different inclinations}

\begin{landscape}
    \begin{figure}
        \centering
    	\includegraphics[width=\columnwidth]{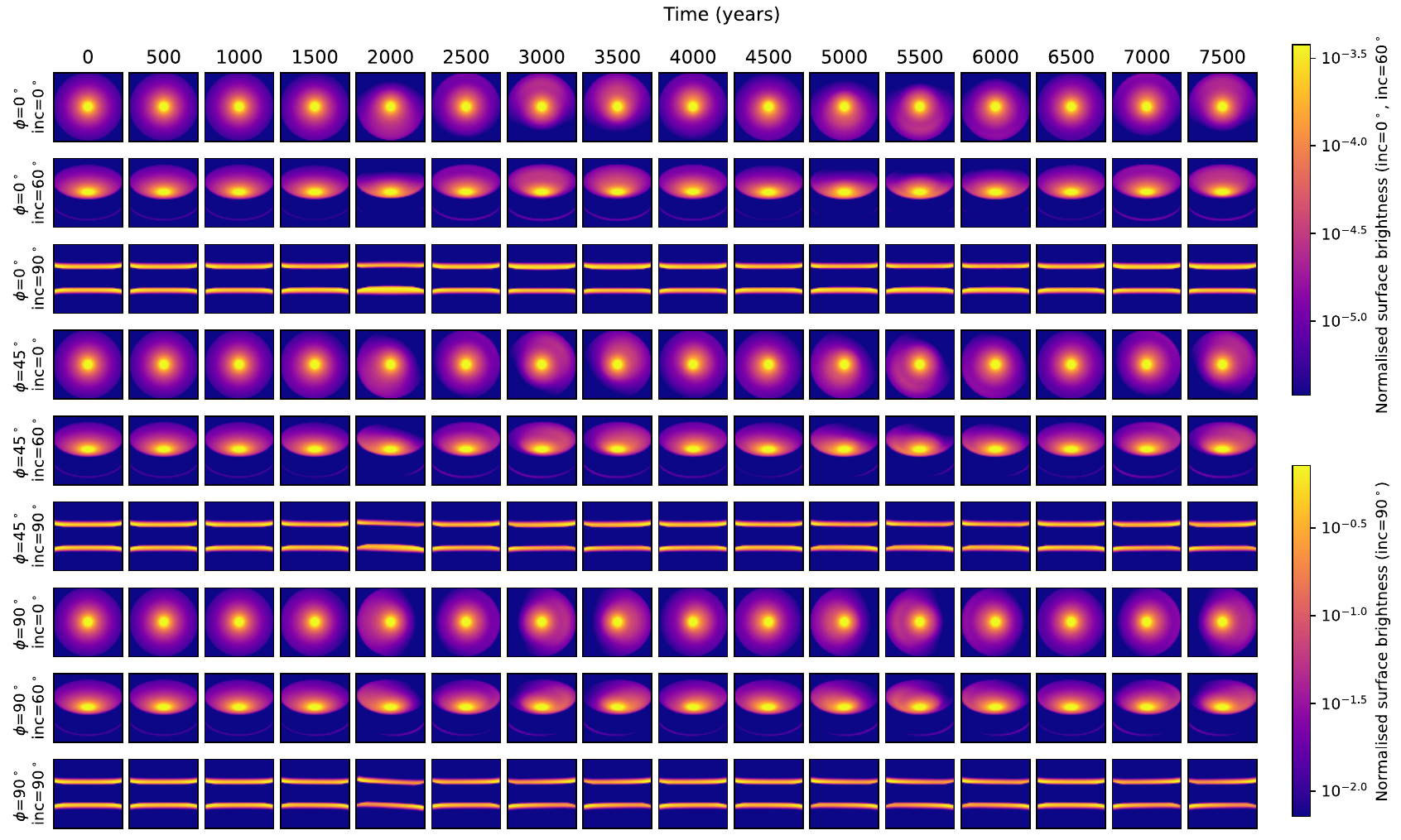}
        \caption{Scattered light images of the shadowing for model 1 for different inclinations of the disc ($inc$) and rotation angles of the disc ($\phi$). Columns show different time steps and rows show the disc for different values of $inc$ and $\phi$. We define $\phi$ to be $0^\circ$ when the shadow is in the north of the disc at $2000\,$years and we rotate clockwise. An inclination of $0^\circ$ is face-on and $90^\circ$ edge-on. The first three rows show $\phi=0^\circ$ for inclinations of 0, 60, and $90^\circ$, the next set of rows are for $\phi=45^\circ$, and the final three for $\phi=90^\circ$. All images are log scaled and normalised to the maximum surface brightness value. We use separate colour bars for $inc$s of 0 and $60^\circ$, and $inc$s of $90^\circ$ so that the variations in brightness are clear for the edge-on discs as they are dimmer (we cannot see the bright central region of the disc).} 
        \label{fig:pa_model1}
    \end{figure}
\end{landscape}

So far in this paper we have focused on shadows seen in face-on protoplanetary discs. However, for most observations of protoplanetary discs the disc has some inclination (statistically it is most likely that the disc is inclined by $60^\circ$ \footnote{The probability of a random angular momentum vector orientation of a disc is uniformly distributed on a sphere so is given by $P\propto\cos(i)$. The median of this distribution is $\cos(i)=0.5$ and so $i=60^\circ$. \label{fn_disc_inc}}). Figure \ref{fig:pa_model1} shows the shadows induced in a disc for model 1 at different inclinations ($inc$) for different rotations of the disc ($\phi$). Columns show different time steps. Each set of three rows shows the shadows for a certain value of $\phi$ for discs inclined by $0^\circ$, $60^\circ$, and $90^\circ$. We define $\phi=0^\circ$ to be where the shadow is in the north of the image at $2000\,$years and we rotate clockwise (so $\phi=90^\circ$ is where the shadow is in the right or west of the image at $2000\,$years). We only present results for shadows with $\phi=0$, 45, and $90^\circ$ because the shadow rocks back and forth and so a $\phi=0^\circ$ will be similar to that of $\phi=180^\circ$ as the shadow will be seen in the north and south of the image for both values of $\phi$.

Shadows are harder to observer at a $60^\circ$ inclination compared with the face-on configuration. For example, for $\phi=0^\circ,\ inc=60^\circ$ (second row of Figure \ref{fig:pa_model1}) we can see a shadow in the North for $2000\,$years, but for $3000\,$years we cannot clearly see a shadow in the South as the flared edge of the disc obscures our view. However, for $\phi=90^\circ$ we do see shadows in the East and West of the disc.

To quantify the visibility of the brightness asymmetries and shadows for different inclinations and $\phi$ values we measure the position of the centre of the flux in x and y for our images. To do this we use the equation:
\begin{align}
    \overline{d} = \frac{\Sigma_{d=0}^{d=n} f_dd}{\Sigma_{d=0}^{d=n} f_d}
	\label{eq: flux_position}
\end{align}
where $\overline{d}$ is the position of the centre of the flux along a certain axis (eg. $x$-axis), $d$ is the position of the flux along a particular axis, $f_d$ is the mean flux at position $d$ (eg. mean flux for each column), and $n$ is the maximum value of $d$. The $x$-axis is along the bottom of an image, and the $y$-axis along the side. This equation calculates the central position of the flux by weighting the position of each pixel ($d$) by the mean brightness of the pixels at that position ($f_d$). This is summed for each position and divided by the total flux. For a perfectly symmetric disc the centre of flux will be the centre of the image, but for a shadowed disc the centre of flux will be skewed away from the shadowed region. This allows us to measure the asymmetry of the images for both models and observations of discs.

\begin{figure*}
	\includegraphics[width=\textwidth]{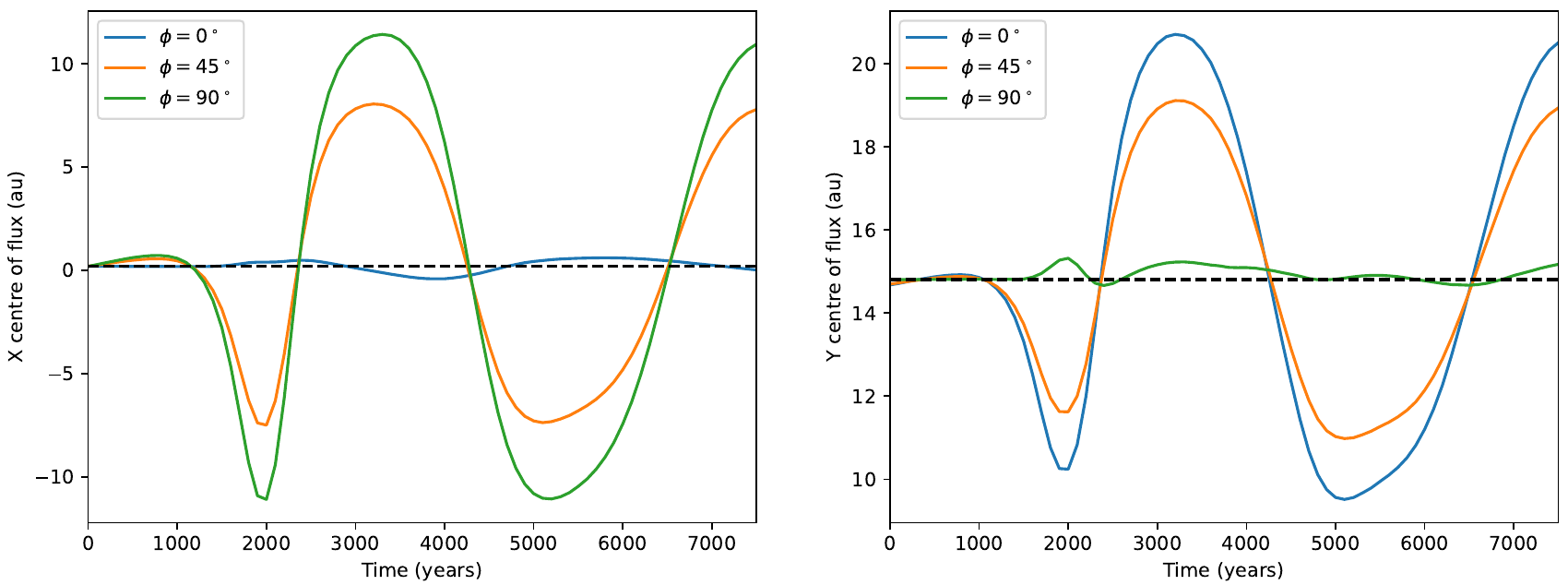}
    \caption{Position of the centre of the flux in x (left) and y (right) with time for images inclined by $60^\circ$. Different colours show different values of $\phi$ (see legend). We can see that for $\phi=90^\circ$ the variation in the centre of the flux is maximum in x and minimum in y. The opposite is true for $\phi=0^\circ$.} 
    \label{fig:xy_inc_comparisons}
\end{figure*}

Figure \ref{fig:xy_inc_comparisons} shows that the centre of flux (for images inclined by $60^\circ$) in $x$ (left) and $y$ (right) changes with time as the shadow moves. We plot the centre of flux for $\phi=0$, 45, and $90^\circ$ (blue, orange, and green lines respectively). For each line plotted we did not include flux values above the $99^{\rm th}$ percentile of the flux (in all images for all time steps at that $\phi$ value). This is because the central part of the images are too bright so no variation in the position of the flux could be measured. 

\begin{figure*}
	\includegraphics[width=\textwidth]{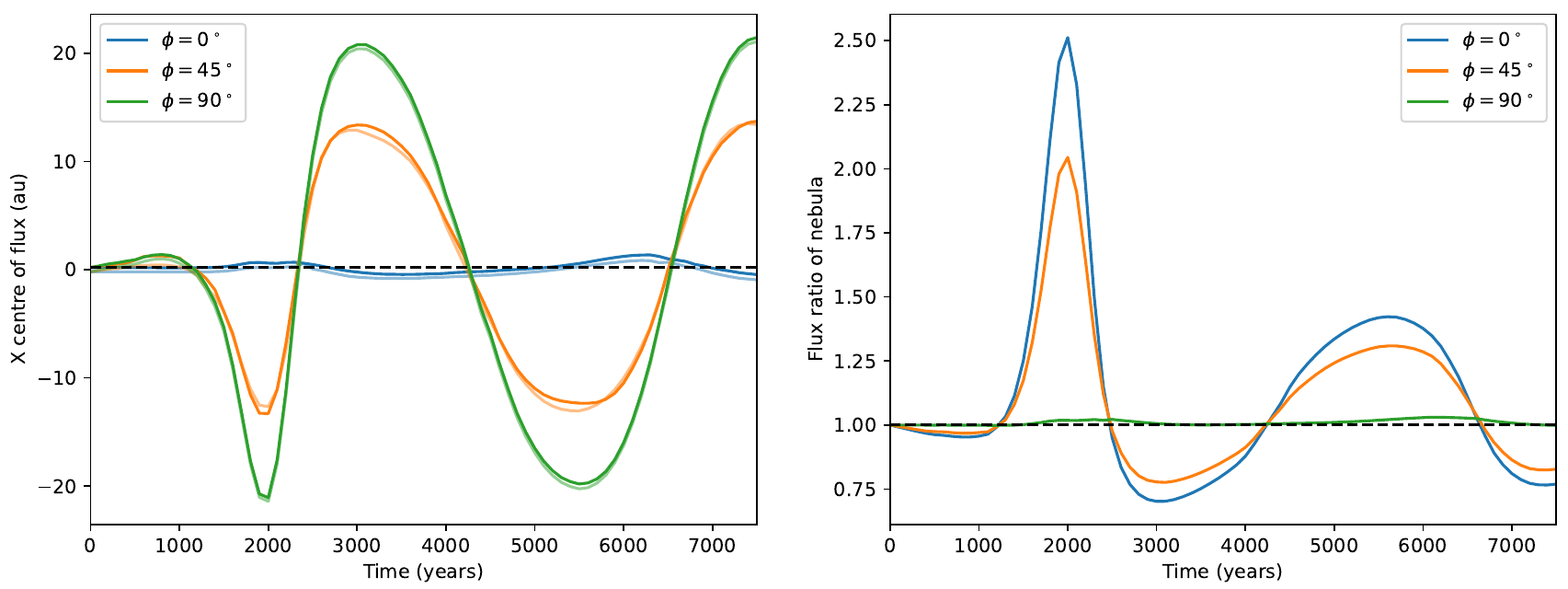}
    \caption{Left: Same as the left of Figure \ref{fig:xy_inc_comparisons} but for the edge-on (inc=$90^\circ$) images for model 1. The different colours show different values of $\phi$ (see legend). The solid lines are for the top nebula of the edge on images and the transparent lines for the bottom nebula flipped across the $y$-axis to show the symmetry of the nebulae. The variation in the centre of the flux is maximum for $\phi=90^\circ$ as this is where the shadow moves from east to west. Right: Ratio of the total flux for the Southern:Northern nebula for different values of $\phi$ (see legend in left image). The flux ratio is greatest for $\phi=0^\circ$ as this is where the shadow moves in the north and south. For $60^\circ$ and $90^\circ$ the ratio peaks at $2000\,$years when the tilt of the disc is the largest.} 
    \label{fig:edge_on_moment}
\end{figure*}

We carried out the same analysis for the edge-on discs as seen in the left of Figure \ref{fig:edge_on_moment}. For the edge-on discs we do not need to exclude the top $99^{\rm th}$ percentile of the flux as the brightest part of the disc is obscured by the midplane (the dark line in the middle of the image). We measure the variation of the central position of the flux in the $x$-axis with time for different rotations of the disc ($\phi=0$, 45, and $90^\circ$ shown by blue, orange, and green lines respectively). We measure the variation for the top nebula (solid line) and the lower nebula (transparent line) which we flip over the $y$-axis to show that there is a symmetry with the top nebula. We also analyse the ratio of the total flux of the Southern:Northern nebula for the edge-on images with different values of $\phi$ (right of Figure \ref{fig:edge_on_moment}).

From these figures we can see that for $\phi=0^\circ$ (blue lines) the asymmetry is greatest in the y-axis (as the shadow is in the North and South of the disc) and there is very little asymmetry in the $x$-axis. The opposite is true for $\phi=90^\circ$ (green lines) as the shadow is in the West and East.

For discs inclined by $90^\circ$ we are seeing light scattered through the edge of the disc, so we do not see shadows, but we do see changes in brightness distribution due to the warp. We see the top and bottom part of the disc (which we call nebula) separated by a dark lane where the scattered light cannot pass through the midplane of the disc. For a perfectly planar edge-on disc the two nebula would be completely symmetric, but for a warped disc we see brightness asymmetries where less light is scattering through the edge of the disc in the region shadowed by the warp \citep[see][for a detailed analysis of the effects of warping on edge-on discs]{villenave_2024, kimming_2025}.

The brightness asymmetries can appear as one nebula brighter than the other \citep[see][]{cuello_2020} (right of Figure \ref{fig:edge_on_moment}) or asymmetries in the nebula themselves - the centre of flux is skewed (left of Figure \ref{fig:edge_on_moment}). The change in brightness ratio between nebula occurs if light is scattering through the edge of the disc from a shadowed region (so there is less light) for one nebula (the darker nebula), but not the other ($\phi=0^\circ$ - row 3 of Figure \ref{fig:pa_model1}). Or, if we rotate the disc $90^\circ$ (so $\phi=90^\circ$ - row 9) we see brightness asymmetries in the nebula themselves because for one half of the nebula light is scattering from a shadowed region (so it is dimmer), but not the other half, so the centre of flux is skewed. We can also see a mix of these effects ($\phi=45^\circ$ - row 6).

However, one nebula can also appear brighter than the other if the disc is not viewed perfectly edge on, due to forward scattering \citep[see][figure 2]{benisty_2023}. For our models the change in the brightness ratio of the two nebulae is partly due to the fact that outer disc tilts (as it is warped), but our viewing angle stays the same. We are therefore no longer viewing the outer disc perfectly edge on. The only way to tell if one nebula being brighter than the other is due to a warp (not the viewing angle) is if we have observations from multiple wavelengths and the brighter nebula switches depending on wavelength of observations \citep[see][]{villenave_2024}. However, it is statistically more likely that the disc is not perfectly edge-on (see footnote \ref{fn_disc_inc}).

To summarise, if the disc is not viewed face-on, or close to face-on, it becomes significantly more difficult to identify shadowing, as the full disc face is not visible. Furthermore, for highly inclined discs (approaching an inclination of $90^\circ$) it is challenging to distinguish whether brightness asymmetries between nebula arise from warping or are simply due to the viewing angle.

\section{Discussion}
\label{sec:discussion}

\subsection{Observational signatures of a flyby in scattered light images}
\label{sec: discussion_signatures}
For scattered light images of a disc with a flyby induced warp we expect to see a broad shadow in the outer disc and an un-shadowed inner disc (with a radius of $41\,$au for model 1). This is because a perturbing flyby induces a warp in the outer disc (casting a broad shadow) which propagates inwards. When it reaches the centre it reflects back, but the central region of the disc with a radius less than the wavelength of the warp remains flat (relative to itself). Therefore the outer disc is shadowed, but not the inner disc.

Since the warp is induced in the outer disc, the warp (and therefore shadow) moves with a low velocity so we do not expect to see the shadow evolve between observations. This is because the warp propagates at half the sound speed of the disc which is given by $c_s(R)=H\Omega$ so, using $H/R$ from Equation \ref{eq: hr_power_law}, $c_s\propto R^{-3/8}$. Therefore a warp propagates more slowly in the outer disc than in the inner disc. We can see this in our results as the oscillation period of the warp for model 1 is $4500\,$years which is too slow for us to see changes between observations in our lifetime. Fast moving shadows, such as TW Hya \citep{Debes_2017, debes_2023} \citep[or variable asymmetries in edge on discs such as in HH 30][]{watson_2007}, are likely due to changes in morphology of the inner disc as a warp in the inner disc propagates more quickly, and changes in the positions of shadows can be seen on shorter timescales. For a flyby though, only the outer disc is warped and so the shadows are not fast moving.

For our flyby models, the speed of the warp is fastest when it reaches the centre of the disc and reflects back so the shadow flips to the other side of the disc. This is when temporal variations of the shadows may be observable if observations were taken decades apart. However, this is when the maximum tilt of the disc is lowest and so the shadow induced is smallest ($\sigma^2$ is lowest) and hardest to observe. The shadow is largest ($\sigma^2$ is highest) when the warp is propagating in the outer disc and the maximum tilt is highest. This is where the shadow is easiest to observe, but detecting variations between observations is very unlikely as the warp is propagating at its slowest and the changes would be very small between observations. Therefore, for a warp induced by a flyby, it is unlikely that changes in the morphology of the disc could be observed over the timescale of decades.

It is worth noting that the period of oscillation of the warps is proportional to the outer radius of the disc: $P\propto R_{\rm out}^{11/8}$ (see Equation \ref{eq: oscillation_period}). Therefore for a smaller disc the period of the warps will be lower and the shadow will move more quickly. For example, for $R_{\rm out}=100\,$au the period of the warp is $\sim 4500$ years and when the shadow is moving fastest (when it switches sides of the disc), the disc spends $\sim 300\,$ years ($7\%$ of $P$) with $\sigma^2>0.01$. Therefore it is unlikely that the switch would be observable. However, for smaller discs where $R_{\rm out}=20\,$au, $P\sim500\,$years and it is likely that when the shadow switches sides $\sigma^2>0.01$ for $\sim35$ years so the shadow could be observed to switch sides of the disc within 40 years. Currently discs of this size (in nearby star forming regions of distances $100-200\,$pc) are not resolvable in scattered light imaging, however disc warps can also be detected from kinematic observations \citep{loomis_2017,perez_2018,Young_2022, winter_2025} taken with ALMA which can potentially probe smaller scales.

We summarise that the characteristics of a warp induced by a flyby are a shadowed outer disc, but an un-shadowed (relatively flat) inner disc and the morphology of the disc is unlikely to change significantly between observations. From our results we have shown that for a low viscosity disc ($\alpha=10^{-4}$) the warp can last for $\sim 10^6\,$years (most of the disc's lifetime) and for $50\%$ of the time the disc has observable shadows. A close flyby encounter may be a common occurrence and so observations of warped discs due to a perturbing flyby may be likely.

\subsection{Do we see shadows which could be due to a flyby-induced warp in observations?}
\label{sec: discussion_observations}

There are multiple examples of discs perturbed by flybys in the literature \citep[see Table 1 and Figure 6 in][and references therein]{cuello_2023}. For example, we see large bridge structures or streamers (eg. HV \& DO Tau \citealt{Winter_2018}, RW Aur \citealt{cabrit_2006, Rodriguez_2018}, Z CMa \citealt{Dong_2022}), and spiral arms (eg. AS 205 \citealt{Kurtovic_2018}, Sag. C cloud \citealt{Lu_2022}). However, these observations do not show obvious shadows due to warps. Of course not every flyby event can cause warps; only sufficiently misaligned flybys provide the necessary torque on the disc \citep{xiang_gruess_2016, nealon_2020}. Alternatively it may be that shadows due to warps are obscured by short-lived larger-scale structures induced by the flyby event.

In many observations of discs perturbed by flybys, where we see large bridge structures or streamers, the flyby was so recent that a candidate perturber has been observed (eg. HV \& DO Tau \citealt{Winter_2018}, RW Aur \citealt{cabrit_2006, Rodriguez_2018}, ZCMa \citealt{Dong_2022}, Sag. C cloud \citealt{Lu_2022}). Close to the flyby event the disc can become extremely perturbed with large scale spirals and mass transfer. For example from hydrodynamic simulations in \citet{cuello_2019} we see that warps are induced in the disc due to the flyby, but so are large scale spirals and bridges (these structures are beyond the scope of the 1D models of this work). These large structures may lift dust to higher latitudes in the disc which may make it more difficult to detect warps in scattered light (see Appendix \ref{appendix:h_effects}). In these cases it may be that close to the flyby event itself other diagnostic tools such as kinematics may be useful for detecting warps (see below).

Also, we emphasise that warps induced by flybys are long lived and have lifetimes on the order of the disc lifetime whereas spirals induced by flybys dissipate over a much shorter time period \citep{kimmig_2026}. Close to the flyby encounter spirals may be more prominent, but flyby induced warps last for a long time after the perturber has passed and are present when the perturber is no longer observable. Observations of warps are often explained by other mechanisms such as a misaligned companion \citep{heap_2000, krist_2005,quillen_2006,nealon_2019}, but these warps could actually be caused by a perturber which can no longer be clearly linked to the system observed. 

We see multiple examples of broad shadows, which are characteristic of flyby-induced warps, in real scattered light observations. For example HD 143006 \citep{Benisty_2018}, WRAY 15-788 \citep{Bohn_2019}, and V1098 Sco \citep{williams_2025}. The favoured explanation for these shadows is a misaligned inner disc, casting a shadow on the outer disc, due to a misaligned perturbing companion rather than a warp induced by a flyby. However, flybys have been proposed as an explanation for misaglined discs \citep[and planets,][]{xiang_gruess_2016} so it is possible that these discs could be misaligned due to a flyby.

Shadows and warps are easiest to detect in face-on discs, but observations of discs are often inclined (see Footnote \ref{fn_disc_inc}). For example \citet{hughes_2009} suggest that the presence of a warp in the disc around GM Aur which may be caused by a stellar flyby. Scattered light observations of GM Aur do not show a shadow in the outer disc \citep{Hornbeck_2016}, however the disc is inclined by $55^\circ$ so a shadow in the outer disc would be harder to detect (see Figure \ref{fig:pa_model1}). 

As mentioned above, the effects of flyby induced warps are more subtle than large scale streamers or spirals and it may simply be that they are harder to detect. Shadows are a direct way to detect warps and from our results we have shown that they are present for a large part of the disc's lifetime ($\tau_\sigma\sim10^5$ for $\alpha=10^{-4}$). However, the mean lifetime of the shadows, $\tau_\sigma$, is less than the mean lifetime of the warp, $\tau$, so shadows may not be visible for more subtle warps. Scattered light images are not the only way to detect warps; \citet{winter_2025} analyse large-scale kinematic structures in protoplanetary discs using ALMA and show that many of the velocity features can be explained by moderate disc warps with tilts of $\sim0.5^\circ-2^\circ$. Therefore for more subtle warps, the warps may be detectable with kinematics, but do not produce shadows in scattered light (see the end of Section \ref{sec:discussion}).

\subsection{How many shadowed discs do we expect to see in clusters?}
\label{sec: discussion_predictions}

From simulations, a close flyby (within $1000\,\rm au$) is likely to occur within a disc's lifetime; \citet{cuello_2023}'s recent review of stellar flybys suggests that for planet-forming discs in a typical stellar environment, >50$\%$ of stars experience a stellar flyby within $1000\,$au. They extrapolate their results from \citet{Pfalzner_2013} who used N-body simulations to model leaky cluster environments and find that for solar mass stars at 1.8 Myr the probability of a flyby event between $100-1000\,$au is $\sim30\%$. 

\citet{Winter_2018} estimate the probability of a flyby event for after 3 Myr for different stellar densities (see their Figure 7). For a close flyby ($<1000\,$au) for stellar densities above 500 $\rm pc^{-3}$ they estimate the probability to be $70\%$ (for stars of 1$M_\odot$ and a velocity dispersion of $4\,$km $\rm s^{-1}$). 

Even for low mass clusters, where flybys were thought to be rare, \citet{pfalzner_2021} have shown that $50\%$ of stars experience a flyby closer than $1500\, \rm au$ as these clusters have a similar density to high-mass clusters. These results suggest that the occurrence of a close ($<1000\,$au) flyby encounter may be common and that the majority of protoplanetary discs may be perturbed by a close flyby at some point in their lifetime.

\begin{figure}
	\includegraphics[width=\columnwidth]{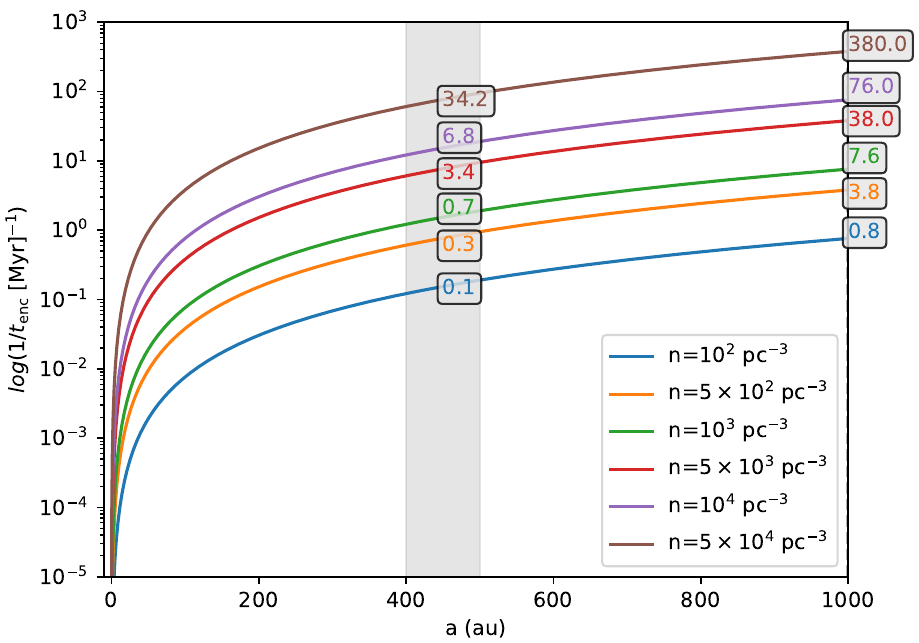}
    \caption{Rate of flyby encounters, per Myr, (calculated using Equation \ref{eq: encounter_rate}) against closest approach for different cluster densities (see legend). The cluster has $\sigma_{\rm d}=4\, \rm km\, s^{-1}$ and $m_{\rm c}=m_{\rm p}=1M_\odot$. Values of the encounter rate are marked for $400<a<500\,$au (grey shaded region) and $a<1000\,$au for different stellar densities (shown by the colours in the legend). We can see that the encounter rate strongly depends on the closest approach of the flyby and the stellar number density of the cluster (note the the y-axis is a log scale).}
    \label{fig:encounter_rate}
\end{figure}

In order to estimate the frequency of observable shadows from flybys in local stellar clusters we first identify the maximum value of $a$ from our results (for a flyby with the most common perturber velocity, $V$, in a cluster) for which shadows due to an encounter would be observable (i.e. $\sigma^2>0.01$). We then calculate the probability of an encounter taking place, with this value of $a$, in stellar clusters of varying densities. From these probabilities we can estimate the expected number of warped/ shadowed discs observable in those environments.

Figure \ref{fig:a_v_tilt_model1} shows the degree of shadowing (quantified by $\sigma^2$) for flyby models with different values of $a$ and $V$ for model 1 (where shadows are observable if $\sigma^2>0.01$). \citet{proszkow_2009} model a wide range of star forming environments and suggest that the mean relative velocity in a stellar cluster is $\sim 3\, \rm km\, s^{-1}$ (see their figure 10). From Figure \ref{fig:a_v_tilt_model1} we can see that for $V=3\, \rm km\, s^{-1}$, the disc is shadowed with $\sigma^2>0.01$ for $a=400$ and $500\,$au (but not for $600\,$au). $\sigma^2$ is still greater than 0.01 after one mean lifetime for these models. For $300\,$au, $V=3\,\rm km\, s^{-1}$ is less than the escape velocity at this radius so models were not run for this velocity. Therefore for $V=3.0\, \rm km\, s^{-1}$ closest approaches of $a=400-500\, \rm au$ produce a shadow which is significant even after one mean lifetime of the warp.

\citet{Winter_2018} calculate the probability of the closest approach of a stellar encounter to be less than a certain distance (after 3 Myr) for different stellar densities (see their Figure 7). In their model the cluster has a uniform density and is made up of stars with mass $1M_\odot$ and velocity dispersion of $4\, \rm km\, s^{-1}$. They calculate the probability that $a<400\, \rm au$, to be $\sim0.3$ for a stellar density of $n=5.0\times10^4\, \rm pc^{-3}$, and $P\sim0.1$ for $n=10^4\, \rm pc^{-3}$. For $a<500\, \rm au$ (which is the maximum value of $a$ for $V=3\, \rm km\, s^{-1}$ which produces significant shadows) the corresponding probabilities are $\sim0.5$ for $n=5.0\times10^4\, \rm pc^{-3}$ and $\sim0.1$ for $n=10^4\, \rm pc^{-3}$. Therefore approximately half of stars in a cluster with a density of $n=5.0\times10^4\, \rm pc^{-3}$ are expected to experience a stellar flyby within $500\, \rm au$ after 3 Myr.

To estimate the number of warped discs in a cluster, we need to calculate the number of flybys which occur in a cluster (of a given density) in a certain time. To do this we can use the encounter rate (number of encounters in a cluster per year), $1/t_{\rm enc}$. This is valid for a star of mass $m_{\rm c}$, travelling in a cluster with uniform density, experiencing perturbations from stars of mass $m_{\rm p}$. In essence, it gives the number of flyby encounters per star per year. We use the equation derived by \citet{lestrade_2011} based on kinetic theory and gravitational focusing:
\begin{equation} \label{eq: encounter_rate}
    \begin{split}
        \frac{1}{t_{\rm enc}}({\rm yr^{-1}}) = 1.9\times 10^{-8}\left( \frac{n}{1000\, \rm pc^{-3}} \right)\left( \frac{\sigma_{\rm d}}{1\, \rm km\, s ^{-1}} \right)\left( \frac{a}{100\, \rm au} \right)^2 +\\
        8.8\times10^{-9} \left( \frac{m_{\rm c}+{m_{\rm p}}}{1\, \rm M_\odot} \right)\left( \frac{n}{1000\, \rm pc^{-3}} \right)\left( \frac{\it a}{100\, \rm au} \right)\left( \frac{1\, \rm km\, s^{-1}}{\sigma_{\rm d}} \right)
    \end{split}
\end{equation}
where $n$ is the stellar number density of the cluster, $\sigma_{\rm d}$ is the velocity dispersion, $a$ is the closest approach of the flyby, and $m_{\rm c}$ and $m_{\rm p}$ are the masses of the central star and perturbing star respectively. The first term is the collision rate calculated from kinetic theory, and the second term accounts for gravitational focusing.

Figure \ref{fig:encounter_rate} presents the encounter rate ($1/t_{\rm enc}$) per Myr as a function of the closest approach of the flyby for different stellar densities (see legend). We assume a velocity dispersion of $4\, \rm km\, s^{-1}$ \citep{Winter_2018} and that the cluster is made up of solar mass stars. Specific values of $1/t_{\rm enc}$ are marked on the graph for $400<a<500\,$au and $a<1000\,$au for different stellar densities. We consider the encounter rate over 1\,Myr as the mean lifetime of a warp in a disc with $\alpha=10^{-4}$ is $\sim1\, \rm Myr$ (see Section \ref{sec:timescale_of_warps}) so if discs were warped by encounters within this time period, we are likely to observe them.

These results show that we expect close flyby encounters to be very common; for $a=400-500\, \rm au$ and $n=5\times10^4\, \rm pc^{-3}$ we expect $\sim34$ flyby encounters to occur per star within $1\,$Myr and $\sim7$ to occur for $n=10^4\, \rm pc^{-3}$. Even for lower stellar densities of $n=5\times10^2\, \rm pc^{-3}$ we would expect $\sim30\%$ of stars to experience an encounter with $400<a<500\,$au (assuming all stars are solar mass stars). We can use Figure \ref{fig:encounter_rate} to estimate the number of encounters (which induce warps with observable shadows) we may expect to see for real clusters.

\subsubsection{Analysis of real clusters}

The Orion Nebula Cluster (ONC) has a high stellar number density of $n=2\times10^4\, \rm pc^{-3}$ \citep{hillenbrand_hartmann_1998} in the inner $0.16-0.21\,$pc and within this cluster lies the Trapezium cluster which has a peak stellar density of $5\times10^4\, \rm pc^{-3}$ \citep{Luhman_2000} (brown line in Figure \ref{fig:encounter_rate}). Therefore $\sim34$ stellar flybys which significantly warp the disc should occur per star per Myr in this cluster. However this cluster is located $\sim410$ pc away \citep{Menten_2007} and so only the largest disc features can be resolved with current scattered light instrumentation. Despite this \citet{valegaurd_2024} present observations of discs in the ONC from SPHERE in which they detect 10 protoplanetary discs, three of which are bright and extended. All three bright discs are asymmetric which they suggest could be due to a warp in the inner disc. These warps could be due to a flyby encounter suggesting that flyby encounters may be common in high density clusters.

$\rm \rho$ Ophiuchus and IC 348 are closer clusters (with distances of $140\,$ pc and $315\,$pc respectively), and have a lower stellar density of $10^3\, \rm pc^{-3}$ (green line in Figure \ref{fig:encounter_rate}) \citep{Luhman_2000}. We expect that over 1 Myr each star experiences $\sim1$ encounter with $400<a<500\,$au. So although the encounter rate is lower than the higher density Trapezium cluster, all discs in this cluster still have the potential to be significantly warped by a flyby over a time period which the warp would be observable.

Although low-density clusters are generally expected to have low encounter rates, this is not always the case. The Taurus star-forming region (located $140$ pc away \citep{kenyon_1994, loinard_2005}) has a low global stellar density of $n=1-10\, \rm pc^{-3}$ \citep{luhman_2000_b} so we may expect flyby encounters to be rare ($1/t_{\rm enc}\sim0.1-1\%\, \rm Myr^{-1}$ for $400<a<500\,$au and $\sigma=4\, \rm km\,s^{-1}$). Nevertheless, \citet{winter_2024} point out that although the stellar density is low, we see evidence of three recent stellar flybys: RW Aurigae \citep{cabrit_2006, Rodriguez_2018}, HV and DO Tau \citep{Winter_2018}, and UX Tau \citep{zapata_2020, menard_2020}. They create a physically motivated model of the Taurus star-forming region including kinematic substructure and binaries and find that stars often experience close encounters and that $\sim25\%$ of discs are tidally truncated (below $30\,$au) due to stellar encounters. This is supported by \citet{Garufi_2024} who carry out a demographical analysis of the Taurus star forming region from VLT/SPHERE observations and find that $30\%$ of discs in their sample are smaller than $30\,$au.

This indicates that stellar flybys may be more common in low density clusters than expected. A possible explanation is that, although the global stellar density is low, Taurus has a structure made up of filaments with 20 high density regions (with a maximum surface density of $\sim2500\, \rm pc^{-2}$) containing $\sim45\%$ of the entire stellar population \citep{gomez_1993, Joncour_2018}. There is evidence that similar high density filaments may be present in other clusters such as $\rm \rho$ Ophiuchus \citep{gomez_1993}.

In a recent paper, \citet{garufi_2026} published a review of scattered light observations of protoplanetary discs from different star forming regions. They find that the percentage of broad shadows observed varies from region to region. In Upper Scorpius and Centaurus $15\%$ and $25\%$ of discs have broad shadows respectively (although only 8 discs were imaged from Centaurus). In Taurus, Ophiuchus, Lupus, and Orion $\sim2.6-5.0\%$ of discs have broad shadows. Chamaeleon, Corona Australis, and $\eta$ Chamaleontis show no discs with broad shadows, although discs in Chamaeleon are faint, discs in Corona Australis are often surrounded by ambient emission, and only 6 discs were imaged from $\eta$ Chamaleontis.

Taking the \citet{garufi_2026} results at face value, the percentage of discs with broad shadows is lower than predicted above, although we note that discs with broad shadows are indeed present in multiple regions. It is worth noting that the observational biases of this sample mean that it is difficult to draw firm conclusions. For example $\sim50\%$ of discs in Orion, Chameleon, and Ophiuchus are non-detections and many of the discs in Chameleon and Orion are faint (see their Figure~3).

\subsubsection{Caveats}

Although these results suggest that the majority of protoplanetary discs should be warped by a flyby encounter, many scattered light observations of protoplanetary discs do not show evidence of warps. Here we outline possible explanations for this discrepancy and implications for protoplanetary discs.

Although we may expect flybys to be a common occurrence, not all flybys cause the disc to warp. For example prograde flybys can cause spirals and tidal stripping, only sufficiently misaligned flybys cause warps \citep{xiang_gruess_2016}. Also, close in flybys can truncate the disc, substantially reducing the outer radius, or even penetrate the disc. Truncated discs are much smaller and so subsequent warps which may be induced may not be observable. Also, since the outer disc radius is reduced a perturber would have to be closer to the disc (than before it was truncated) to induce a warp. Similarly, in our models we set the outer disc radius to be $100\,$au, but smaller discs would require a closer flyby encounter to perturb the disc which is less likely to occur. For example, for a disc with a radius of $20\,$au the truncation radius would be $\sim60\,$au and the rate of flyby encounters (for $r=60\,$au, $n=5\times10^4\rm \,pc^{-3}$) is only $\sim1.4\,\rm star^{-1}\, Myr^{-1}$ compared to $34.2\,\rm star^{-1}\, Myr^{-1}$ for $r=300\,$au (see Figure \ref{fig:encounter_rate}).

Furthermore, we calculate the encounter rate assuming stellar mass encounters (as this is what we use for our models), but high mass encounters may also be common and are likely to have a stronger, potentially more destructive, impact on the disc. We can see this from Equation \ref{eq: flyby_torque} which gives the torque due to the perturber; the torque is directly proportional to the perturber's mass so the higher the mass the higher the impact on the disc. Conversely, lower mass encounters, or encounters where the perturber has a higher velocity may not have a big enough impact on the disc to perturb it.

The encounter rate is high for clusters with a high stellar density, but in less dense clusters, such as the Taurus star-forming region, there are still regions where stars may form in relative isolation (separate to the high density filaments). In these regions, where $n\sim1-10\,\rm pc^{-3}$, the encounter rate is much lower ($\sim 0.1-1\%\, \rm Myr^{-1} $) and we would not expect to see many discs warped due to flyby encounters.

If we do not observe many warped discs in higher density regions, this may have implications for the viscosity of protoplanetary discs. In higher density clusters, such as $\rho$ Ophiuchus (which is close enough for discs to be observed), where $n\sim 10^3\,\rm pc^{-3}$, we expect about one stellar mass encounter per star (with $400<a<500\,$au and $\sigma=4\,\rm km\,s^{-1}$). Thus we may expect to observe many warped discs in this region. If we do not then this is an important constraint on the nature of the viscosity in protoplanetary discs. One possibility is that the viscosity parameter is higher than $\alpha = 10^{-4}$, which would reduce the time to damp the warp. \citet{ogilvie_2013} have suggested that a warp might lead to hydrodynamic instability in the disc that could lead to a larger $\alpha$ in a warped protoplanetary disc than in an un-warped one (see also \citealt{drewes_2012}). For example, if $\alpha = 10^{-3}$ then the lifetime of the warp is only $10^5$ years and the likelihood of observing a significantly warped disc is reduced explaining the low number of observations of warped discs.

Additionally, we do not take into account multiplicity in our analysis. Most stars form in binaries (or multiple stellar systems: see \citealt{offner_2023}) which may affect the structure of the disc. Close binaries (with separations of a few hundred au or less) have a significant effect on the disc \citep[eg.][]{terquem_1993,paploizou_terquem_1995} and may even tidally truncate the disc \citep{artymowicz_1994}. However for wider binaries the perturbations due to flybys may be more significant than the companion. Also, the companion itself may have a similar effect to a flyby if it has a very large separation ($>1000\,$au). \\\\ From our analysis we conclude that flybys are a common occurrence in stellar clusters. For example, stars in high density clusters such as the Trapezium cluster, which has a stellar density of $5\times10^3\, \rm pc^{-3}$, experience $\sim34$ encounters per Myr (for stellar mass encounters with $400<a<500\,$au and $\sigma=4\, \rm km\, s^{-1}$). For $\rho$ Ophiuchus and IC 348, which have $n=10^3\, \rm pc^{-3}$, stars experience one flyby per Myr. The lifetime of a flyby induced warp (for $\alpha=10^{-4}$) is about a Myr so warped discs would likely be observable in these systems. Even for clusters with lower stellar densities observations suggest that fly-by encounters are common which has been attributed to higher density regions in these clusters.

We have shown that shadows due to flyby induced warps are long lived and likely observable in scattered light. Systematic scattered light surveys of clusters with different densities searching for large scale warps should help determine if flyby induced warps are common and constrain $\alpha$ in protoplanetary discs. If warped discs due to flyby encounters are not observed as regularly as expected this may tell us that $\alpha>10^{-4}$ in protoplanetary discs and so the warps are damped more rapidly. 

Warps can also be detected using high resolution ALMA observations to look for kinematic signatures of warps \citep[eg.][]{loomis_2017, perez_2018, Young_2022, winter_2025}. Kinematic observations may be able to detect more subtle warps for which shadows are unobservable. In fact \citet{winter_2025} suggest that many of the observed large-scale velocity features in discs can be explained by warps and that warps may be common. In the future we will adapt our models in combination with molecular line radiative transfer to predict the observational kinematic signatures of the warps.

Also, if new instrumentation facilitates scattered light observations in smaller discs, this could reveal temporal variations of shadows due to flyby induced warps (as the period of oscillation of the warp is proportional to the outer radius of the disc so temporal variations are visible over shorter timescales in smaller discs).

\section{Summary}
We have modelled the propagation of warps in protoplanetary discs due to a perturbing flyby using 1D warp propagation theory. We then used a fast radiative transfer code to model scattered light images of the disc and the shadows cast by the warp. We model flybys with parabolic trajectories orthogonal to the disc plane and at an angle to the disc plane. For both models a warp wave is induced in the outer disc which propagates back and forth through the disc long after the flyby has passed. Flybys induce different oscillations with a combination of different twists and tilts. This casts a broad shadow in the outer disc which can oscillate back and forth or rotate.

We have shown that the likelihood of warps being excited by flybys is high and that the induced warps have long lasting observable consequences in scattered light. This suggests that flyby induced warps are common which is in agreement with recent kinematic observations by \citet{winter_2025} who suggest that many large scale velocity features in discs can be explained by warps.

Our main conclusions are:
\begin{itemize}
  \item A protoplanetary disc can become significantly warped due to a flyby encounter. The outer disc warps (inducing a shadow in the scattered light images) and this warp oscillates back and forth (causing the shadow to rock back and forth in the outer disc or to rotate). The inner disc remains un-warped within a radius less than the wavelength of the warp ($41\,$au for $a = 300\, \rm au,\ V=V_{\rm esc}$). 
  \item It is likely that shadowing from the warp due to a flyby would be observable for most of the disc's lifetime. For a low viscosity disc ($\alpha=10^{-4}$) perturbed by a flyby with a closest approach of $300\, \rm au$ (travelling at $V=V_{\rm esc}$) the lifetime of the warp is $\sim10^6\,$years and is on the order of the lifetime of the disc itself. Therefore the disc would be significantly warped for most of its lifetime. We have also shown that when the shadow oscillates back and forth, for $50\%$ of the time (in one mean lifetime of the warp) the disc is significantly shadowed (as the disc has a maximum azimuthal variance above $0.01^\circ$). Therefore it is likely that the disc would be shadowed for most of its lifetime.
  \item The period of oscillation of the warp is proportional to the outer radius of the disc and so we may only see shadows evolve between observations in small discs (for $R_{\rm out}\sim20\,$au we may see the shadow switch sides of the disc in $\sim40\,$years). Small discs are currently difficult to observe in scattered light, but may be observable with future instrumentation. In larger discs the warp propagates slowly so we are unlikely to see shadows evolve between observations in our lifetime because the warp is induced in the outer disc and travels at $c_s/2$.
  \item We vary the closest approach ($a=300,\ 400,\ 500,\ 600\, \rm au$) and velocity ($V=2.44 - 6.88\, \rm km\, s^{-1}$) of the flyby and show that for a velocity of $3\, \rm km\, s^{-1}$ flybys with $a=400-500\, \rm au$ can significantly warp the disc so shadows are observable (i.e. $\sigma^2>0.01$). We estimate that flyby encounters are common in high density clusters: for clusters with $n=5\times10^4\, \rm pc^{-3}$ we expect to see $\sim34$ flyby encounters per star per Myr with $400<a<500\, \rm au$ (and $\sigma=4\, \rm km\, s^{-1}$) which could warp the disc enough for shadows to be observable. If observations suggest that warped discs are not as abundant as our estimates this may imply that the warps are damped more rapidly and $\alpha>10^{-4}$ in protoplanetary discs.
\end{itemize}

Overall we conclude that flybys can induce a warp in the outer regions of the disc, resulting in a broad shadow in scattered light. This warp is long lasting so shadows can be observed even if the flyby encounter took place much earlier in the disc's lifetime the perturbing body is no longer detectable.

\section*{Acknowledgements}
We warmly thank Matthew Bate for useful discussions. KLM thanks Daniel Price for hosting her at Monash University for three months where some of this work was carried out and for useful discussions. We thank the anonymous referee for their time and for providing detailed and useful feedback. This work was supported by STFC studentship no. 2697199 and the STFC funded long-term attachment. CJN acknowledges support from the Leverhulme Trust (grant No. RPG-2021-380). This work made use of NUMPY \citep{harris_2020}, MATPLOTLIB \citep{Hunter_2007}, SCIPY \citep{virtanen_2020}, and PLOTLY \citep{plotly}.

%%%%%%%%%%%%%%%%%%%%%%%%%%%%%%%%%%%%%%%%%%%%%%%%%%
\section*{Data Availability}
The 1D warp modelling code used to carry out this work is available via Zenodo (\url{https://doi.org/10.5281/zenodo.17700556}). Data is available on reasonable request to the authors.



\bibliographystyle{mnras}
\bibliography{bib} % if your bibtex file is called example.bib

%%%%%%%%%%%%%%%%%%%%%%%%%%%%%%%%%%%%%%%%%%%%%%%%%%

%%%%%%%%%%%%%%%%% APPENDICES %%%%%%%%%%%%%%%%%%%%%

\appendix
\section{Benchmark tests}
\label{appendix:a}
In this appendix we present the results of the benchmark tests carried out to check that our 1D warp code could reproduce previous models.

\subsection{Evolution of an arbitrary warp}
\label{sec:appendix_martin}
\begin{figure}
    \centering
	\includegraphics[width=\columnwidth]{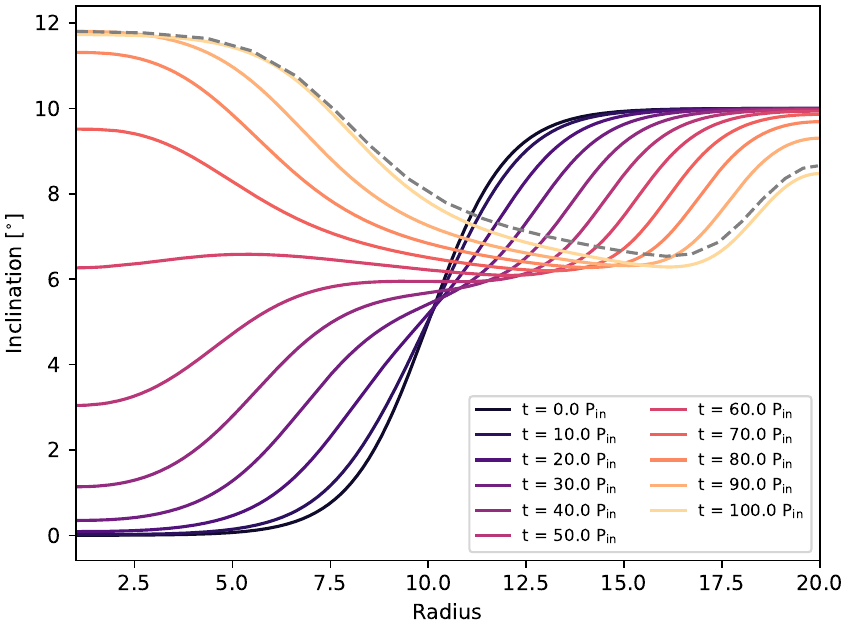}
    \caption{Figure showing the evolution of an arbitrary warp in the wave-like regime with no external torque. The inclination of the disc is plotted against radius. Different colours show different times, given by integer values of the time period of the inner disc ($P_{\rm in}$), as shown by the legend. The black line, at $t=0$, shows the initially warped disc which flattens out over time. These results are reproduced from \citet{Martin_2019} and match their Figure 1. The final time step from their Figure 1 has been manually extracted and plotted as a grey dashed line.}
    \label{fig:inc_vs_r_arbitrary_warp}
\end{figure}

Firstly we tested our code with no external torque ($\textbf{\textit{T}}=0$) by setting up an arbitrary warp in the wave-like regime and allowing it to evolve with time. We set up our disc as outlined in Section 4 of \citet{Martin_2019} where the initial inclination of the disc is given by a $\tanh$ function; the inclination is $0^\circ$ at the inner edge, $10^\circ$ at the outer edge, and the disc is warped at $r=10\,$au. 

Figure \ref{fig:inc_vs_r_arbitrary_warp} shows how this warp changes with time. The inclination of the disc is plotted against the radius of the disc ($R_{\rm in}=1$, $R_{\rm out}=20$). Different times, given by rotations of the inner disc ($P_{\rm in}$), are plotted in different colours as shown by the legend. The initially warped disc is shown by the black line ($t=0$) and the warp starts to flatted out over time. This model reproduces the results from \citet{Martin_2019} as it matches the results in their Figure 1. The results from \citet{Martin_2019} have been manually extracted and plotted for the final time step ($t=100\,P_{\rm in}$) as a grey dashed line.

\subsection{Warp induced by a rotating black hole}
\label{sec:appendix_lubow}
\begin{figure}
    \centering
	\includegraphics[width=\columnwidth]{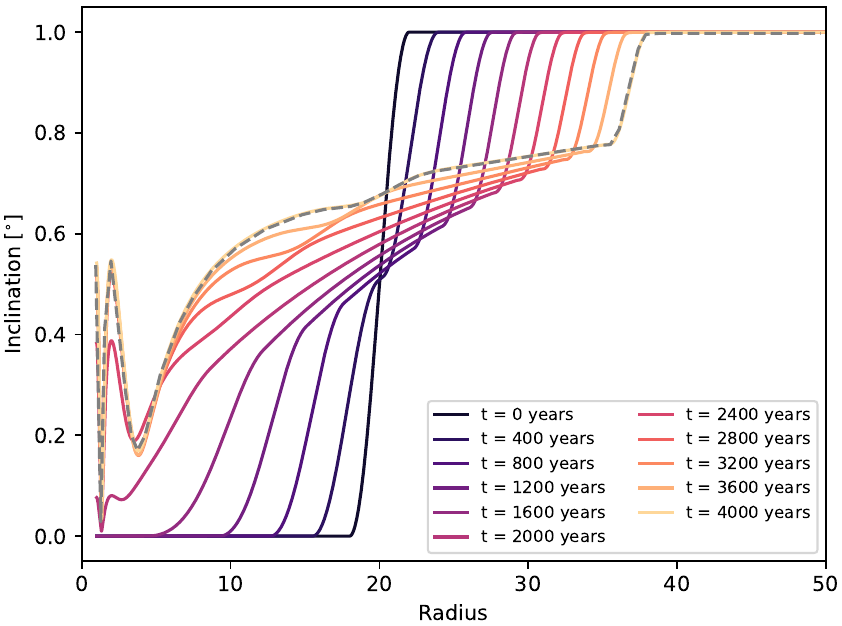}
    \caption{Figure showing the evolution of a warp induced by a black hole due to Lense-Thirring precession. The inclination of the disc is plotted against radius with times shown by different colours as shown by the legend. The disc has an initial inclination given by a sinusoidal step function (black line). These results are reproduced from \citet{lubow_2002} and match their Figure 3. The grey dashed line shows the results from \citet{lubow_2002} which have been manually extracted and plotted for the final time step.}
    \label{fig:inc_vs_r_lubow_2002}
\end{figure}

Here we present the results of a warp induced by a rotating black hole due to Lense-Thirring precession taking into account the apsidal and nodal precession which occurs due to the black hole. We reproduce the model outlined in section 4 of \citet{lubow_2002}. The disc has an initial inclination given by a sinusoidal step function.

Figure \ref{fig:inc_vs_r_lubow_2002} shows the results of our model match the results from \citet{lubow_2002} (see their figure 3). The inclination of the disc is plotted against radius with different times shown by different colours (see legend). The final time step ($t=4000\,$years) has been manually extracted from \citet{lubow_2002} and plotted as a grey dashed line.

\subsection{Warp induced by a flyby orthogonal to the disc plane}
\label{sec:appendix_np10}
\begin{figure}
    \centering
	\includegraphics[height=0.875\textheight]{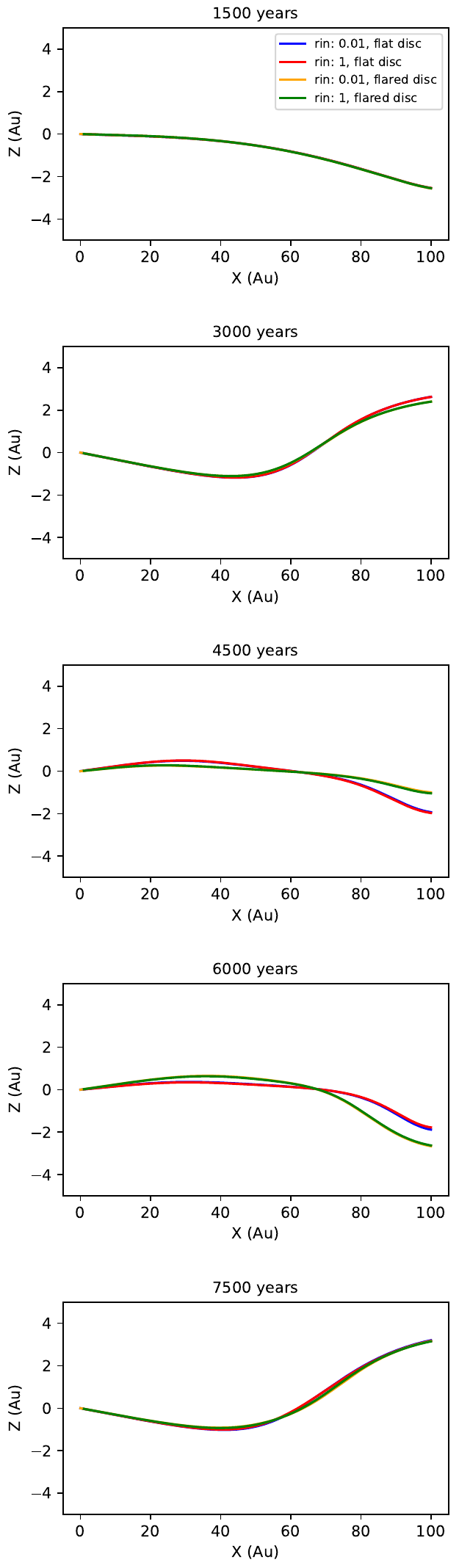}
    \caption{Figure showing how the disc is warped at different time steps. The figure shows $z$ position against $x$ position (effectively the height of the midplane against radius) for model 1. We show models for $H/R=0.1$ for $R_{\rm in}=0.01$au (blue) and $R_{\rm in}=1$au (red) to show that there is very little difference between these models. We also show models for a flared disc for $R_{\rm in}=0.01$au (orange) and $R_{\rm in}=1$au (green), again showing little difference between them.} 
    \label{fig:zpos_model1}
\end{figure}

Here we present the results of the warp induced by the flyby described in model 1 of \citet{Nixon_Pringle_2010}. We set-up the model as described in \citet{Nixon_Pringle_2010} (see Section \ref{sec: flyby_models}) with a constant $H/R=0.1$. Figure \ref{fig:zpos_model1} shows how the warp propagates through the disc with time. It shows the $z$-position of the disc for at different x-positions. This is a effectively a cross-section of the disc showing the height of the disc at each radius. Since the disc warps about the $x$-axis (as the flyby is travelling parallel to the $z$-axis with y=0) this cross-section allows us to see the full warp structure. The disc is symmetric so we do not need to show the negative x-direction.

We can see the shape of the disc from our results match that in \cite{Nixon_Pringle_2010} as Figure \ref{fig:zpos_model1} matches their Figure 1. However, we note that the tilt of this disc is approximately half that as in \citet{Nixon_Pringle_2010} as they are missing a factor of two in their approximation of the torque.

The blue line in Figure \ref{fig:zpos_model1} (under the red line) shows the model with an inner radius of $0.01\,$au (which matches the model in \cite{Nixon_Pringle_2010}), and the red line shows the model for an inner radius of $1\,$au. We can see very little difference between the models with different inner radii. We also plot our models for the flared disc (see Section \ref{sec: flyby_models}) with an inner radius of $0.01\,$au (orange) and $1\,$au (green). Again, there is no significant difference. Therefore for the modelling in this paper we use an inner radius of $1\,$au as this is computationally less expensive.

\section{Effect of scale height on shadows seen in scattered light images}
\label{appendix:h_effects}

\begin{figure*}
	\includegraphics[width=\textwidth]{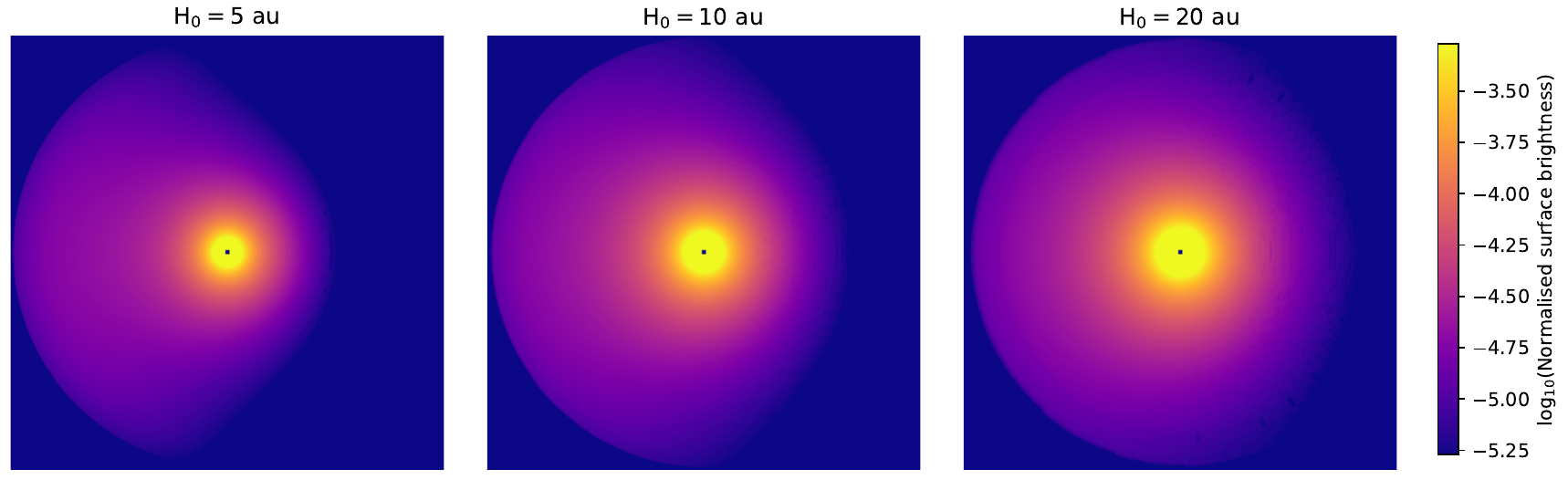}
    \caption{Scattered light images for model 1 at $t=2000\,$years for different values of $H_0$. The snapshots were taken at $2000\,$ years, when the flyby reaches its closest approach, as this is when the disc is most warped (see Section \ref{sec:results}) and a shadow is clearly visible in the outer disc. The left images shows the flattest disc with $H_0=5\, \rm au$, for the middle image $H_0=10\, \rm au$, and the right image shows the most flared disc with $H_0=20\, \rm au$. Each image is normalised to the maximum surface brightness value and log scaled.}
    \label{fig:hr_comparison}
\end{figure*}

\begin{figure}
	\includegraphics[width=\columnwidth]{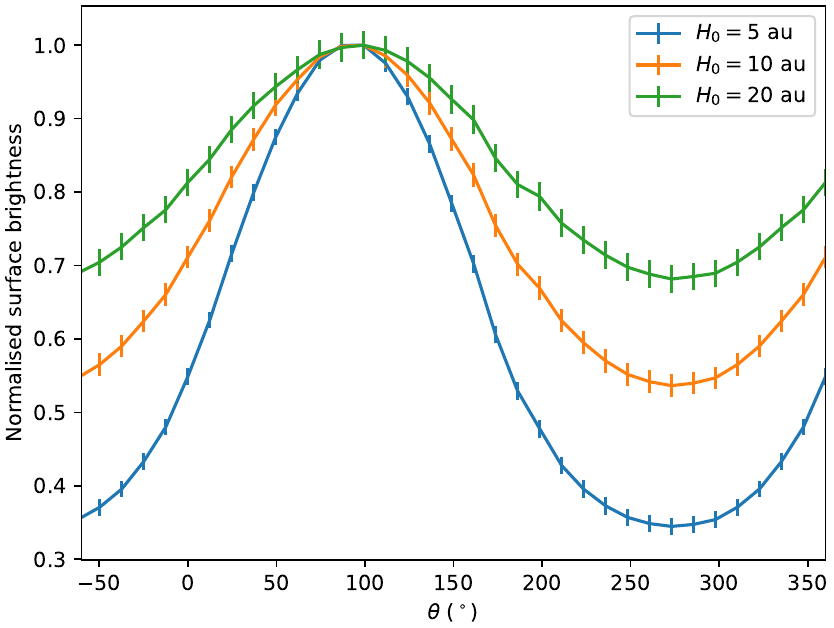}
    \caption{Azimuthal profiles of surface brightness for the images in Figure \ref{fig:hr_comparison} with $H_0=5$, 10, and $20\,$au (blue, orange, and green lines respectively). Each profile taken between $10-100\,$au with 30 azimuthal bins and is normalised to the maximum surface brightness of the disc. $\theta=0^\circ$ at the top of the image and increases moving anticlockwise round. The dip in the azimuthal profile due to the shadow is deepest (with respect to the maximum value) for the flattest disc (with $H_0=5\,$au). Also the shadow extends the furthest azimuthally for the flattest disc. For the most flared disc (with $H_0=20\,$au) the shadow has the lowest surface brightness (with respect to the brightest value) and extends the least far azimuthally. This can also be seen in the images in Figure \ref{fig:hr_comparison}.}
    \label{fig:hr_azimuthal_profiles}
\end{figure}

It is beyond the scope of the paper to perform an exhaustive investigation of disc parameters and their impact on the shadows, and we instead chose to focus on a canonical, representative, disk. Nonetheless we are aware that our results will be sensitive to some disc parameters, particularly the disc scale height and we performed some test calculations to illustrate this.  Figure \ref{fig:hr_comparison} shows scattered light models for the disc (for model 1 at $t=2000\,$years) for $H_0=5$, 10, and $20\,$au (left, middle, and right respectively). The images are normalised to their maximum surface brightness and log scaled. The flattest disc, with $H_0=5\,$au is the most shadowed as more of the outer disc is blocked from the star due to the warp. The most flared disc, with $H_0=20\,$au is the least shadowed, although there is still a clear shadow in the outer disc.

Figure \ref{fig:hr_azimuthal_profiles} shows azimuthal profiles of these scattered light images with $H_0=5$, 10, and $20\,$au shown in blue, orange, and green respectively. Each profile is taken between $10-100\,$au and normalised to the maximum surface brightness value. The shadow is deepest (compared to the maximum surface brightness) and broadest for the flattest disc ($H_0=5\,$au). The shadow is shallowest and extends the least (azimuthally) for the most flared disc ($H_0=20\,$au). This can also be seen in the scattered light images. For this paper we choose a scale height with $H_0=10\,$au, but it is worth bearing in mind that a flatter disc would be more shadowed and a more flared disc, less shadowed.

\section{Comparison of our fast radiative transfer code with full radiative transfer}
\label{appendix:fast_rt_vs_full}

\begin{figure*}
	\includegraphics[width=\textwidth]{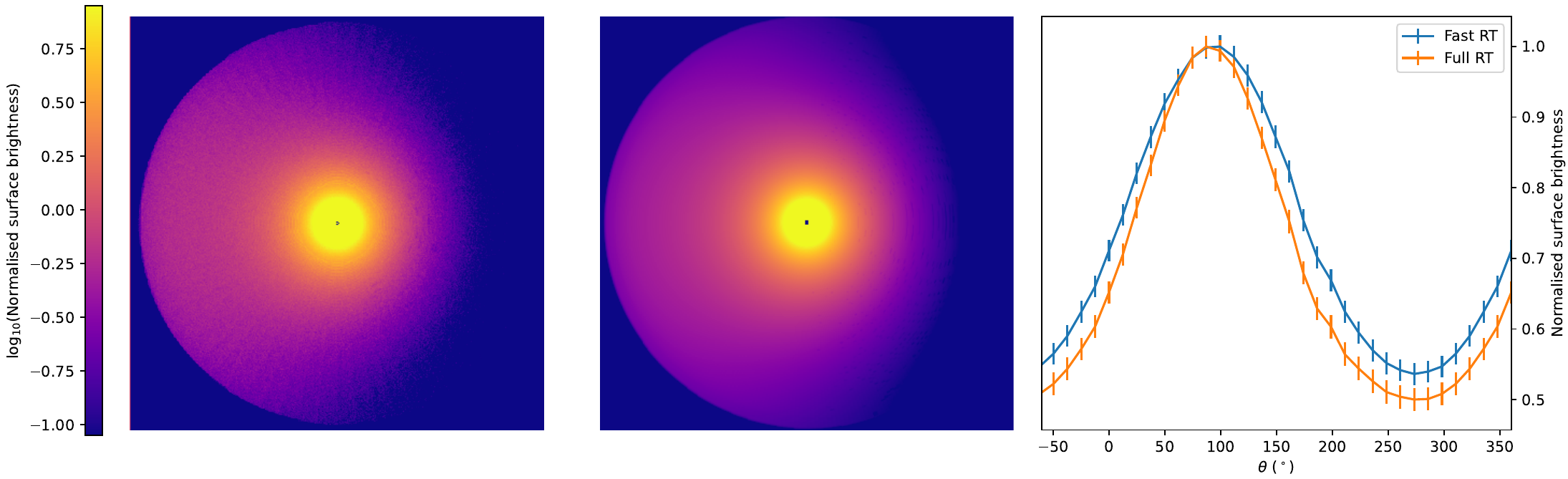}
    \caption{Comparison of scattered light images created using the full radiative transfer code, TORUS \citep{Harries_2019}, (left) and our fast radiative transfer code (middle). These images are of model 1 at $t=2000\,$years (when the flyby reaches its closest approach) as this is when the disc is most warped (see Section \ref{sec:results}) and a shadow is clearly visible in the outer disc. Both images are normalised to the $90^{\rm th}$ percentile of their surface brightness values and log scaled. On the right we show azimuthal profiles of the surface brightness for the fast (blue) and full (orange) radiative transfer codes. The profiles are taken between $10-100\,$au and normalised to the maximum surface brightness value of that profile. $\theta=0^\circ$ at the top of the image and increases moving anticlockwise round. We can see that the width and depth (relative to the brightest point in the disc) of the shadow are very similar for both codes and that the fast radiative transfer code is a good approximation of the full code for this work.}
    \label{fig:rt_tor_comparison}
\end{figure*}

In order to confirm that the simplifying assumptions that underpin our fast RT code are valid, we set up our fiducial disc model in the Monte Carlo radiative transfer code TORUS \citep{Harries_2019}. We computed radiative equilibrium to get the dust temperatures in the disc, before calculating a $J$-band $512 \times 512$ pixel scattered light image that includes both the near-IR thermal emission as well as the direct stellar radiation, adopting $10^9$ photon packets to reduce the Monte Carlo noise.

Figure \ref{fig:rt_tor_comparison} shows that our fast radiative transfer code reproduces shadows produced with the TORUS radiative transfer code very well. We show scattered light images from a snapshot of model 1 at $t=2000\,$years (the closest approach of the flyby) produced using TORUS (left) and the fast radiative transfer code (right). Images were taken at this time as this is when the disc is most warped for this model and so a large shadow is visible in the outer disc (see Section \ref{sec:results}). The images were normalised to the $90^{\rm th}$ percentile of the surface brightness values and log scaled. Both images clearly show a broad shadow in the outer disc and an un-shadowed inner disc.

From the rightmost image of Figure \ref{fig:rt_tor_comparison} we can see the extent as to how well the scattered light images match for the full and fast radiative transfer codes. The plot shows the azimuthal profiles of the surface brightness from the scattered light images produced using the full radiative transfer code (orange) and the fast radiative transfer code (blue). Each profile is taken between $10-100\,$au and normalised to the maximum surface brightness value. $\theta=0^\circ$ in the North of the images and increases moving anticlockwise round. The shadow has almost the same width and depth (compared to the brightest part of the disc) for both radiative transfer codes. For the full code the shadow is slightly deeper and broader (azimuthally), but overall the shape is very similar. Therefore our fast radiative transfer code is a good approximation of the shadows seen in scattered light due to warps for the models in this paper.

\section{Measurements of \texorpdfstring{$\sigma^2\ $}\ examples}
\label{appendix:b}

\begin{figure}
    \centering
	\includegraphics[height=0.87\textheight]{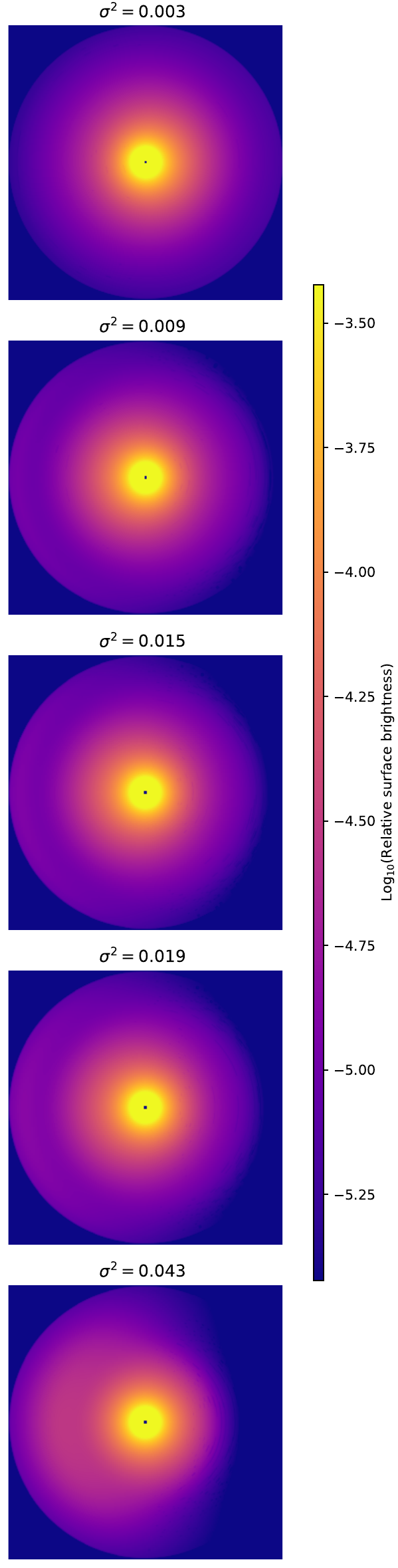}
    \caption{Figure showing scattered light images with shadows of model 1 for different values of $\sigma^2$. All images are normalised to the maximum surface brightness value of the images. Shadows with $\sigma^2$ greater than 0.01 are observable. The top image has $\sigma^2\ll0.01$ so the shadow is unobservable, below $\sigma^2\lesssim0.01$ so although we can see a small shadow on the right it is unlikely to be observable. The three images below these show observable shadows with increasing values of $\sigma^2$}
    \label{fig:sigma2_examples}
\end{figure}

Here we present examples of scattered light images of warped discs with different values of $\sigma^2$ for model 1 (see Figure \ref{fig:sigma2_examples}). $\sigma^2$ measures the azimuthal asymmetry of the surface brightness of scattered light images and gives us a measure of how shadowed the disc is (see Section \ref{sec:asymmetry_parameter}). If $\sigma^2>0.01$ then the disc is significantly shadowed.

The top image in Figure \ref{fig:sigma2_examples} shows a disc with a very low azimuthal asymmetry with $\sigma^2=0.003$. This is much less than the threshold value for a shadow to be observable of 0.01 and no shadow is visible. The next image down shows a shadow with $\sigma^2=0.009$. This is close to the threshold value and a shadow is visible in the right, but it is unobservable. In the image below this we see an observable shadow (on the right) with $\sigma^2=0.015$. The bottom two images show shadows with increasing values of $\sigma^2$. We can see that the deeper the shadow, the more azimuthally asymmetric the disc is, and the higher the value of $\sigma^2$.

%%%%%%%%%%%%%%%%%%%%%%%%%%%%%%%%%%%%%%%%%%%%%%%%%%

% Don't change these lines
\bsp	% typesetting comment
\label{lastpage}
\end{document}